\documentclass[sigconf]{acmart} 

\usepackage{multirow}
\usepackage{tabularx}

\AtBeginDocument{%
  \providecommand\BibTeX{{%
    \normalfont B\kern-0.5em{\scshape i\kern-0.25em b}\kern-0.8em\TeX}}}

\copyrightyear{2021} 
\acmYear{2021} 
\setcopyright{acmcopyright}\acmConference[CHI '21]{CHI Conference on Human Factors in Computing Systems}{May 8--13, 2021}{Yokohama, Japan}
\acmBooktitle{CHI Conference on Human Factors in Computing Systems (CHI '21), May 8--13, 2021, Yokohama, Japan}
\acmPrice{15.00}
\acmDOI{10.1145/3411764.3445294}
\acmISBN{978-1-4503-8096-6/21/05}

\begin{document}

\title{A Visual Analytics Approach to Facilitate the Proctoring of Online Exams}





\author{Haotian Li}
\affiliation{%
  \institution{Department of Computer Science and Engineering, HKUST, Hong Kong SAR, China}
}
\email{haotian.li@connect.ust.hk}

\author{Min Xu}
\affiliation{%
  \institution{Department of Computer Science and Engineering, HKUST, Hong Kong SAR, China}
}
\email{mxuar@connect.ust.hk}

\author{Yong Wang}
\affiliation{%
  \institution{School of Information Systems, Singapore Management University, Singapore}
}
\email{yongwang@smu.edu.sg}

\author{Huan Wei}
\affiliation{%
  \institution{Department of Computer Science and Engineering, HKUST, Hong Kong SAR, China}
}
\email{hweiad@connect.ust.hk}

\author{Huamin Qu}
\affiliation{%
  \institution{Department of Computer Science and Engineering, HKUST, Hong Kong SAR, China}
}
\email{huamin@cse.ust.hk}
\renewcommand{\shortauthors}{Li~\textit{et al.}}
\newcommand{\haotian}[1]{\textcolor{black}{#1}}
\newcommand{\haotiancomment}[1]{\textcolor{red}{[haotian: #1]}}

\begin{abstract}
Online exams have become widely used to evaluate students' performance in mastering knowledge in recent years, especially during the pandemic of COVID-19. 
However, it is challenging to conduct proctoring for online exams due to the lack of face-to-face interaction.
Also, prior research has shown that online exams are more vulnerable to various cheating behaviors, which can damage their credibility.
This paper presents a novel visual analytics approach to facilitate the proctoring of online exams by analyzing the exam video records and mouse movement data of each student.
Specifically, we detect and visualize suspected head and mouse movements of students in three levels of detail, which provides course instructors and teachers with convenient, efficient and reliable proctoring for online exams.
Our extensive evaluations, including usage scenarios, a carefully-designed user study and expert interviews, demonstrate the effectiveness and usability of our 
approach.
\end{abstract}





\maketitle

\vspace{-0.5em}
\section{Introduction}


With the rapid development of online learning in the past decade, online exams and tests are becoming increasingly popular for course instructors to assess the knowledge of students~\cite{costagliola2009visualization}. 
For example, Massive Open Online Courses (MOOCs) such as Coursera and EdX often require students to pass a series of online exams before they can gain a final course certificate. Meanwhile, conventional universities also continue to expand their online course programs and hold online exams for students~\cite{michael2013student}. Such a trend is further significantly accelerated from 2019 due to the COVID-19 lockdown, and most schools and universities have switched to embrace online teaching and online exams.
However, one major challenge for online exams is \textit{how to proctor online exams in a convenient, efficient and reliable manner}.
Prior research~\cite{prince2009comparisons,richardson2013strengthening,rogers2006faculty,king2014cheating} has shown that online exams are vulnerable to cheating behaviors.
According to the survey by King and Case~\cite{king2014cheating}, about $74\%$ of students in 2013 reported that it is easy to cheat in online exams and nearly $29\%$ of the students indicated that they cheated in online exams.
These cheating behaviors can damage the credibility of online exams, which makes online exam proctoring crucial for MOOCs platforms and universities to further expand the application and usage of online exams.


Different from traditional exams with onsite proctoring, online exams lack face-to-face interactions. It brings trouble to the proctoring of online exams and various types of cheating behaviors may occur in online exams~\cite{ullah2016classification}. 
For example, students may type the questions into the browser and search for possible solutions from the Internet. They may also send messages to a third party (e.g., friends) to ask for help by using their mobile phones or chat apps on the computer. 
Without face-to-face interactions in online exams, it is not an easy task to identify such cheating behaviors.
To enable effective proctoring, existing online exams usually ask the students to use webcams
to monitor and record their activities during the exams~\cite{atoum2017automated,holden2020academic,moten2013examining,prathish2016face_detection_head_pose,li2015moop}. Accordingly, a set of preliminary studies on the proctoring of online exams have been conducted based on such kinds of settings.





According to our survey, the existing approaches for the proctoring of online exams can be generally categorized into three groups: manual proctoring, fully automated proctoring, and semi-automated proctoring.
Manual proctoring is commonly applied in the proctoring of online exams and many online testing solutions, such as Kryterion\footnote{\href{https://www.kryteriononline.com/}{https://www.kryteriononline.com/}} and Loyalist Exam Services\footnote{\href{http://loyalistexamservices.com/}{http://loyalistexamservices.com/}}, employ such proctoring. Specifically, it requires a few proctors watching the videos of all the students during the whole online exam, which is often labour-intensive and time-consuming.
Instead, fully automated proctoring aims to reduce the manual efforts of proctors by utilizing machine learning techniques to analyze the recorded video and audio data of students during the online exam~\cite{chuang2017timedelay,Roth2014typing,Roth2015typing}. It automatically detects suspected behaviors and classifies them into cheating or non-cheating.
However, it is often difficult for the existing fully-automated proctoring methods to achieve a very high accuracy and the validation of the results becomes an issue. 
To mitigate this issue, some recent online exam proctoring approaches combine the detection by machine learning approaches and further manual confirmation by proctors~\cite{li2015moop, gosia2018human}. 
But their manual confirmation relies on manually checking the original videos backwards and forwards, which is still inconvenient and time-consuming for proctors.




In this paper, we propose a novel visual analytics approach to facilitate the proctoring of online exams.
Inspired by prior studies, our approach combines human efforts with machine learning techniques to achieve \textit{convenient, efficient and reliable proctoring} for online exams.
Specifically, our approach analyzes both the exam videos of each individual student recorded by a webcam and the mouse movement data.
To collect the mouse movement data, we design and implement a lightweight JavaScript plugin that can be easily embedded into different web pages including common web-based learning management systems (e.g., Canvas\footnote{\href{https://www.instructure.com/canvas/}{https://www.instructure.com/canvas/}}) and does not require students to add extra settings. 
With the collected videos and mouse movement data, key features indicating suspected exam cheating behaviors, including both abnormal head movements (e.g., abnormal head rotation, face disappearance from the screen)
and mouse movements (e.g., copy and paste, moving the mouse out of the exam web page),
are extracted.
Furthermore, we designed effective visualizations to enable the interactive exploration of student cheating behaviors in three levels of detail: \textit{Student List View} provides an overview of the cheating behaviors of all the students through a list of radar-chart-based glyphs; \textit{Question List View} visualizes the cheating risk distribution of all the questions finished by each student and \textit{Behavior View}, along with \textit{Playback View},
enables the detailed inspection of a student's suspected cheating behavior distribution of working on a specific question and its comparison with other students and questions.
Compared with prior proctoring approaches that need multiple extra devices or sensors~\cite{atoum2017automated,li2015moop}, 
students are only required to have one webcam on their computer, which is often available for most laptops and makes our approach \textit{convenient} to deploy in real online exams. Also, our straightforward and effective visualizations help users \textit{efficiently} investigate the student cheating behaviors of different levels, and the detailed comparisons across different students and questions enable a more \textit{reliable} cheating behavior judgement. 
We extensively evaluated the effectiveness and usability of our approach through three usage scenarios, a user study and expert interviews.
The major contributions of this paper can be summarized as follows:

\begin{itemize}
    \item We formulate the design requirements for the proctoring of online exams by working together with domain experts (i.e., university faculty and teaching staff) and surveying prior studies.
    \item We propose a novel visual analytics approach for the proctoring of online exams by visualizing the head and mouse movements of students during online exams in three levels of detail, enabling convenient, efficient and reliable proctoring.
    \item We conduct extensive evaluations, including three usage scenarios, a carefully-designed user study and expert interviews, to demonstrate the effectiveness and usability of the proposed approach. 
\end{itemize}








\section{Related Work}
The related work of our paper can be categorized into three parts: online proctoring methods, mouse movement visualization and head pose analysis.

\subsection{Online Proctoring}
Online exams are emerging nowadays with the population of online learning. 
The methods of online proctoring can be categorized into three types: online human proctoring, semi-automated proctoring and fully automated proctoring. 
Online human proctoring means that there will be remote proctors watching students during the whole online exam. It is a very common method used by many online testing solution providers (e.g., Kryterion, Loyalist Exam Services) and some universities (e.g., University of Amsterdam)~\cite{gosia2018human}.
However, it is very labor-intensive and the cost will be high when a large number of students attend an online exam.

To eliminate the usage of manpower, some fully automated proctoring approaches are proposed~\cite{atoum2017automated, chuang2017timedelay}, which often use machine learning techniques to identify cheating behaviors. 
Currently, there are some other online proctoring platforms, including ProctorU\footnote{\href{https://www.proctoru.com/}{https://www.proctoru.com/}} and Proctorio\footnote{\href{https://proctorio.com/}{https://proctorio.com/}}, using 
automated proctoring based on machine learning.
However, all the existing fully automated proctoring approaches suffer from similar concerns as other machine learning methods in education. 
These concerns include the ``black box'' nature of the machine learning algorithms and unreliable decision making led by biased training datasets~\cite{southgate_blackmore_pieschl_grimes_mcguire_smithers_2019}.
Due to these concerns, it is almost impossible to totally rely on automated methods to determine whether a student cheats in an online exam or not.

To address the problem resulting from the fully automated proctoring methods, semi-automated proctoring has been introduced to involve humans in the final decision making~\cite{li2015moop, gosia2018human, costagliola2009visualization}. One representative prior work is Massive Open Online Proctor proposed by Li~\textit{et al.}~\cite{li2015moop}. 
Specifically, their approach first detects suspected student cheating behaviors with machine learning techniques and the detection results will be further checked by teachers.
However,
it does not provide teachers or instructors with a convenient way to explore and analyze suspected student cheating behaviors.
Also, it requires that each student in the online exam uses multiple devices (e.g., two webcams, a gaze tracker and an electroencephalogram~(EEG) sensor) to record their exam process, which is not affordable for most educational institutions. 
Migut~\textit{et al.}~\cite{gosia2018human} proposed a method to calculate the similarity between two successive frames in videos which record screens and extract video clips with dissimilar frames for manual checking. 
Their method also suffers the problem that there is no convenient method to explore the students' behaviors in the extracted video clips. Futhermore, detecting cheating behaviors using local materials (e.g., paper materials, mobile phones) is not supported in their method.
Costagliola~\textit{et al.}~\cite{costagliola2009visualization} proposed a visual analytics system to assist teachers in invigilating an exam. However, it is limited to detecting the cheating case that a student is looking at another student's screen, which is not common in online exams.

Inspired by the prior research above, we aim to propose a visual analytics system for the efficient detection and analysis of various common suspected cheating behaviors in online exams. It will utilize easily-collected data and combine the domain knowledge of users with machine computation power.


\subsection{Mouse Movement Visualization}

Mouse movements are commonly used to analyze user behaviors and cope with various tasks including user modeling~\cite{mueller2001user_modeling, liu2015mouse_website}, cognitive load evaluation~\cite{grimes2015cognitive, khawaji2014cognitive}
and student performance prediction~\cite{wei2020mouse_student, li2020mouse_student}. 
Raw mouse movement data is spatial-temporal data and hard for humans to interpret.


A few visualization approaches have been proposed to visualize the spatial and temporal information of mouse movement data, including 2D and 3D visualization.
2D visualizations often plot the spatial information on a vertical axis and a horizontal axis and encode temporal information in a weaker visual channel~(e.g., colors).
Arroyo~\textit{et al.}~\cite{arroyo2006mouse_visualization} plotted raw mouse trajectories on web pages and used a heatmap-like design to show the time delay on each element in the website. The occlusion of mouse trajectories is severe in their method when the trajectory is complex.
Burigat~\textit{et al.}~\cite{burigat2008mobile} implemented a 2D visualization to draw all mouse movement trajectories on web pages and use colors of lines to encode the sequential information of movements. This method also suffers from the problem of occlusion and it is difficult to track the sequential order of movements.
Heatmap is a frequently used technique in 2D visualization of mouse movement data. 
The frequency of mouse movement data in an area is represented by the colors in the heatmaps. 
Currently, several web analytics tools including Hotjar\footnote{\href{https://www.hotjar.com/}{https://www.hotjar.com/}} and Mouseflow\footnote{\href{https://mouseflow.com/}{https://mouseflow.com/}} apply heatmaps to present the mouse movement data.
The drawback of this method is that they cannot show any detailed movement and the temporal information is also lost.
Region-of-Interest (ROI)-based visualization has also been explored by prior studies~\cite{xia2019mouse_student, xia2020qlens, brown2014user_modeling}.
They visualized the transitions between ROIs to conduct visual analysis of user mouse movement behaviors. However, these methods depend highly on the appropriate definitions and choices of ROIs.
3D visualizations have also been proposed to present mouse movement data.
Zgonnikov~\textit{et al.}~\cite{zgonnikov2017mouse_visualization} proposed a landscape-like design to visualize the positions and speeds of mouse movements. 
Leiva and Viv{\'o}~\cite{leiva20123d} plotted line charts in three dimensions to represent the mouse movement positions and temporal information, respectively.
However, they share common limitations with other 3D visualization methods, including occlusion and inaccurate depth perception.

In this paper, we propose a novel visual design for showing mouse movements to support the visual analytics of students' suspected cheating behaviors during online exams.
%

\subsection{Head Pose Analysis}
Head pose estimation is an important topic in the research on computer vision. 
Head poses show how the head rotates in three dimensions (i.e., yaw, pitch, roll) which are illustrated in Figure~\ref{fig:head_pose}.
The representative methods on head pose estimation include FSA-Net~\cite{yang2019fsanet}, PADACO~\cite{kuhnke2019padaco} and Hopenet~\cite{ruiz2018hopenet}.

Head pose estimation has also been applied in the proctoring of online exams. For example, Prathish~\textit{et al.}~\cite{prathish2016face_detection_head_pose} applied a head pose estimation method proposed by Narayanan~\textit{et al.}~\cite{narayanan2014headpose} to detect misconduct.  
Chuang~\textit{et al.}~\cite{chuang2017timedelay} extracted students' head poses with the method proposed by Baltrusaitis~\textit{et al.}~\cite{Baltrusaitis2012headpose}. 
They calculated several statistical metrics of head poses (e.g., average yaw angle, maximum pitch angle), 
which were further input into their regression models to predict whether a student had cheated or not in the exam.



In this paper, we also consider analyzing and further visualizing head pose data, which help proctors conduct a convenient and reliable exploration of cheating behaviors in online exams.

\section{Requirement Analysis}
\label{requirement_analysis}
To better understand the
major challenges and design requirements in conducting proctoring for online exams, we have worked closely with five teachers (university professors or teaching staff) (P1-P5) at our university in the past six months.
P1 is a professor who has taught several online courses on human computer interactions and data analytics in past years and has also been working on research projects on E-learning for more than five years.
P2 is an assistant professor who has rich teaching experience and has also taught multiple online courses on programming languages. 
P3 and P4 are full-time teaching associates who are mainly responsible for assisting professors on proctoring exams and marking papers. 
P5 is a lecturer who has instructed multiple online courses on design and innovation.
All experts have experience in organizing and proctoring online exams.
P1 is also a co-author of this paper.
We conducted a series of interviews and discussions with them through online video meetings and email communications.
We collected their feedback and summarized the major
design requirements for proctoring online exams.
\haotian{
We denote the five major design requirements as \textbf{R1}-\textbf{R5} for easy reference in the subsequent sections.
}


\textbf{R1. Identify students at high risk of cheating.}
According to the feedback of our experts,
all of them agreed that it is almost impossible to manually review all the videos, since 
there are often multiple or even several hundred students taking the same online exam for a course. 
Therefore, an effective approach to facilitate the proctoring of online exams should help teachers or other proctors easily and quickly identify the students who have possibly cheated in the online exam, especially those students of high risk.
This is also the fundamental goal of any approach for enabling the efficient proctoring of online exams.



\textbf{R2. Locate the questions where
high-risk cheating behaviors occurred.} 
When a student is identified as at high risk of cheating during an online exam,
it is often necessary for teachers and other proctors to further explore where and when the student has cheated and check how he/she cheated.
However, the current way to achieve this is to manually go through the original videos, which is often time-consuming.
For example, P3 commented that
\textit{``A typical online exam lasts for 2-3 hours and may involve hundreds of students. Reviewing individual people is time-consuming''}. P2 also described this method as \textit{``a dull process but needs 120\% concentration''}, which increases the burden on teachers.
To handle this problem, 
it is important to
locate the questions a student may cheat on and enable a fast review and check of cheating behaviors.

\textbf{R3. Inspect students' detailed
cheating behaviors.
} 
All the experts agreed that they also need to inspect detailed cheating behaviors to better understand the detected suspected cases.
According to the feedback of experts, there are various cheating methods including using unauthorized paper materials, seeking help through social media and searching for answers on the Internet. Among them, the majority of commonly-seen cheating methods are related to head and mouse movements.
For example, a sign of cheating on other web pages is that the student's mouse arrives at the edge of the exam web page and stays for a while, as P3 suggested. Also, P4 pointed out that turning the head to somewhere else also can indicate some cheating behaviors such as using cheat sheets.
Thus, teachers need to inspect the detailed mouse and head movements during the online exam.
For such kind of inspections, a convenient and intuitive way to explore those behaviors, which can indicate cheating, is highly appreciated.  




\textbf{R4. Confirm suspected cheating cases through comparison across students and questions.}
P1 suggested that cheating cases are always hard to confirm, since some normal behaviors also look like cheating such as rotating the head to read the questions.
Thus, our system needs to provide a convenient and effective approach to confirm that a suspected case is not led by normal behaviors.
P4 agreed that comparison with peers is an important way to avoid the systematic errors led by question design and personal habits when reviewing suspected cases. 
In practice, different question designs may lead to different problem-solving behaviors of students during the online exam. For example, long questions on the screen may require students to rotate their heads to read them. Also, students' habits may affect their behaviors during the online exam. 
Thus, a comparison with peers' behaviors on the same question and a student's own behaviors on other questions can help teachers and proctors reduce the possibility of making mistakes in judging a suspected cheating case.

\textbf{R5. Explore the original video and mouse movement data in a convenient manner.}
As P2 and P3 suggested, a functionality of playing back video recordings and mouse movements is essential. It can help proctors to further confirm suspected cheating cases, which may be wrongly labeled by some automated detection methods. 
For example, drinking water could be easily recognized as abnormal behavior, since students move their heads severely to drink water. Also, due to the unstable network or errors of webcams, one or two frames of the video may be of low quality, which can result in face detection failure.
Besides, proctors may be interested in finding if there are any other suspected cases in the video that may not be detected by the current approaches.
\section{Data Collection}
\label{section:data_collection}

To enable the convenient proctoring of online exams,
we propose using the video taken by the front webcam and mouse movement data during the online exams to detect cheating behaviors. 
Since most laptops have a webcam at the front, it is convenient to use it to record a student's behavior, such as head movements, during an exam. 
Mouse movement data is also collected, as mouse movements
reflect where a student is focusing and are easier to be collected without extra devices than other means such as eye tracking~\cite{costagliola2009visualization, chen2001mouse_eyetracking, cooke2006mouse}. 
Prior studies have also explored other features for cheating detection, such as audio~\cite{atoum2017automated}, eye movements and electroencephalogram (EEG)~\cite{li2015moop}.
However, our approach does not include those, 
as they are not always available in real online exams. 
For example, as our experts P1 and P3 suggested, audio is often unavailable in online exams held through online meeting software, since students are usually muted to avoid noise.
\haotian{Furthermore, the collection of eye movement and EEG data requires extra devices, which bring more costs and limit their usage in practice.}
In this section, we introduce the details of our data collection.



\subsection{Data Collection Set-up}
\label{section:data_collection_setup}

\haotian{There is no available public data for the proctoring of online exams. Therefore, we decided to hold a mock online exam to collect data after a careful discussion with P1.
The major reason for collecting data from a mock exam is that we can ask participants to indicate where and how they cheated. Such information can be used as the ground truth for evaluating the effectiveness of the proposed approach. However, it is difficult to ask students to indicate where and how they have cheated in real online exams.
To ensure that the mock online exam has a similar setting as that of real online exams, we worked closely with P1 and designed an online exam consisting of two question sets. In the mock online exam, participants' detailed exam records are collected, including their exam videos, mouse movement data, duration of the online exam, grades and the exact labeling on their cheating behaviors.} 

\textbf{Exam.}
We designed a mock online exam consisting of two question sets with both sets focusing on evaluating students' knowledge of JavaScript. Each question set consists of 10 multiple choice questions and 4 short answer questions. 
For example, a multiple choice question can look like this: \textit{``Which value will not be returned by the `typeof' in JavaScript? A. number; B. object; C. function; D. null''} and a short answer question will ask students to list at least 3 ways to empty an Array in JavaScript.
The time limit for one question set was 25 minutes and participants were allowed to submit early after he/she finished all the questions.

\textbf{Participants and Apparatus.}
In our mock online exam, we recruited 24 participants (7 female, $age_{mean} = 24.75$, $age_{sd} = 1.94$) \haotian{by social media}. 
They are all postgraduate students or fresh graduates who have experience in using JavaScript.
They received US \$2.5 if they finished the whole online exam. Also, to encourage participants to take the online exam seriously and act as taking a real exam, US \$0.375 was paid for each correctly answered question.


To mimic the environment of real online exams, \haotian{as suggested by P1,} the mock online exam was conducted online. We implemented a web-based online exam system to collect webcam video recordings and mouse movement data. Participants were asked to show their entire faces in the video recordings. Before the online exam, we sent each participant an exam guideline and a cheatsheet on key JavaScript knowledge points that are useful for both question sets.

\textbf{Procedure.}
\haotian{Before the mock online exam, we first introduced the exam guideline and clarified what data would be collected in the exam. Then we asked for their permission to allow us to use collected data for research purposes.
}
In the mock online exam, each participant needed to finish both question sets and was asked to cheat on one of them. 
To eliminate the influence of the difference between two question sets,
we arranged the participants and the order of the question sets required for cheating in a counterbalanced manner.
In the question set on which they were asked to cheat, participants needed to use at least 3 methods to cheat \haotian{and on each question, they could only apply one cheating method. 
The questions to cheat on and the detailed cheating methods were decided by participants.
Additionally, in our instruction, we emphasized that they needed to try their best to pretend they were in a real closed-book online exam.}
When participants finished the question set on which they were asked to cheat, a questionnaire will appear and ask them to indicate where and in which way they cheated.
Participants were also allowed to have a break for 5 minutes between two question sets. Since early submissions are permitted, their exact time used to answer each question set is also recorded as the duration between their entrance to the exam web page and their final submission.
In the following week, we graded all the exam scores of each participant.



\subsection{Collected Data}
\label{sec:collected_data}
Table~\ref{table:collected_data} shows the basic statistics of our collected data.


\begin{table}[h]
\caption{Statistics of collected data.}
\Description[This table shows the statistics of collected data.]{In our mock online exams, we collected 48 videos whose total length is about 9.5 hours. We also collected 286,940 mouse movement records and 239 cheating cases in the local environment or on the computer.
}
\centering
\small
\setlength{\aboverulesep}{0.5pt}
\setlength{\belowrulesep}{0.5pt}
\vspace{-1em}
\begin{tabular}{l|l|l}
\toprule
\multicolumn{2}{l|}{Number of participants} & 24   \\ \hline
\multicolumn{2}{l|}{Number of videos} & 48 \\ \hline
\multicolumn{2}{l|}{Number of mouse movement records}  & 286,940 \\ \hline
\multicolumn{2}{l|}{Length of videos} & 9h 21m 32s \\\hline
\multirow{2}{*}{Number of cheating cases} & In the local environment & 50 \\ \cline{2-3}
& On the computer & 189 \\\bottomrule

\end{tabular}
\vspace{-1em}
\label{table:collected_data}
\end{table}


\textbf{Video Data.}
In our mock online exam, we collected 48 videos in total, which are 30 frames per second (FPS) with a resolution of $640\times480$.
The time length of a video varies from 8 minutes to 20 minutes, as participants were allowed to conduct early submission if they had finished all the questions. Since videos are taken by the front webcams, we can learn how students move their heads in these videos and detect if they have cheated.

\textbf{Mouse Movement Data.}
In our online exam, students used their mouses and keyboards to interact on the exam web page to answer questions. 
\haotian{Since most of the interactions are conducted by mouse movements, we use the term \textit{mouse movement data} to denote all the collected interaction data generated by mouse or keyboard interactions.}
We developed a JavaScript plugin\footnote{The plugin is available at  \href{https://github.com/HKUST-VISLab/Mousetrack}{https://github.com/HKUST-VISLab/Mousetrack}.} to collect mouse movement data, which is implemented based on the DOM structure of the HTML file and is generalizable to collect mouse movements on any other web pages. Furthermore, the plugin is automatically loaded with the web page and does not require proctors or students to conduct any extra settings. 
Since the plugin only works on one web page and cannot collect any data after the student finishes the online exam, it will not lead to 
the privacy concerns of collecting data in the background.

In each record of the mouse movement data, the mouse position and the DOM event type\footnote{\href{https://www.w3schools.com/jsref/dom\_obj\_event.asp}{https://www.w3schools.com/jsref/dom\_obj\_event.asp}} are recorded. We collected six types of DOM events:
\begin{itemize}
    \item \textit{Blur}: the web page loses focus, which is triggered when the participant leaves the current web page. 
    \item \textit{Focus}: the web page is the current focus, which is triggered when the participant enters the web page.
    \item \textit{Copy}: content is copied from the current web page \haotian{by using either mouse or keyboard.}
    \item \textit{Paste}: content is pasted to the current web page \haotian{by using either mouse or keyboard.}
    \item \textit{Mousemove}: mouse moves on the web page.
    \item \textit{Mousewheel}: mouse wheel rolls to scroll on the web page.
\end{itemize}
Among all the six types, \textit{``blur''}, \textit{``focus''}, \textit{``copy''} and \textit{``paste''} are considered as indicators of cheating, since their occurrence always reflect the usage of some external materials on the computer\haotian{, as indicated by P3.}
\haotian{However, our expert P4 also pointed out that it is possible that a student may copy and paste some materials merely on the exam web page. 
Thus, individual ``copy'' or ``paste'' is insufficient to judge if some cheating behaviors are occurring.
To address this issue, we can check if ``blur'' and ``focus'' exist in the context of ``copy'' and ``paste'' mouse events to verify suspected cheating behaviors. 
If ``blur'' and ``focus'' exist around ``copy'' and ``paste'', the student may copy some questions, use unauthorized materials and paste something to the exam web page. 
For example, in Scenario 3 of Section~\ref{usage_scenario}, the student copied something from the exam page and left the exam page for a while to cheat by running the copied code.
If there is no ``blur'' or ``focus'' around ``copy'' and ``paste'', the student is likely just copying and pasting on the exam web page and the possibility of cheating is low.
}


\textbf{Cheating Types.} 
\label{section:cheating_types}
By analyzing our collected data, we classify students' cheating methods into two types:
\begin{itemize}
    \item \textit{Cheating in the local environment}: students use unauthorized materials locally to cheat, for example, paper materials and mobile phones, while they stay on the exam web page. The common feature of these cheating methods is that students need to turn their head away from the current screen. Thus, we propose using \textit{face disappearance} and \textit{abnormal head pose} as indicators of cheating in the local environment.
    \item \textit{Cheating on the computer}: students leave the exam web page and use the computer to access unauthorized materials, for example, searching on the Internet, asking friends through social media and using electronic notes. 
    Also, students always copy the questions to search or paste answers to the website to save time. 
    Thus, leaving the website and ``copy and paste'' are our main indicators for finding suspected cheating behaviors on computers.
\end{itemize}

\section{Our Approach}
In this section, we introduce the technical details of our visual analytics approach\footnote{Our system is available at \href{https://github.com/HKUST-VISLab/Visual-analytics-approach-online-proctoring}{https://github.com/HKUST-VISLab/Visual-analytics-approach-online-proctoring}.} for the proctoring of online exams. 

\subsection{System Overview}

\begin{figure}[hbt!]
    \centering
    \includegraphics[width=\linewidth]{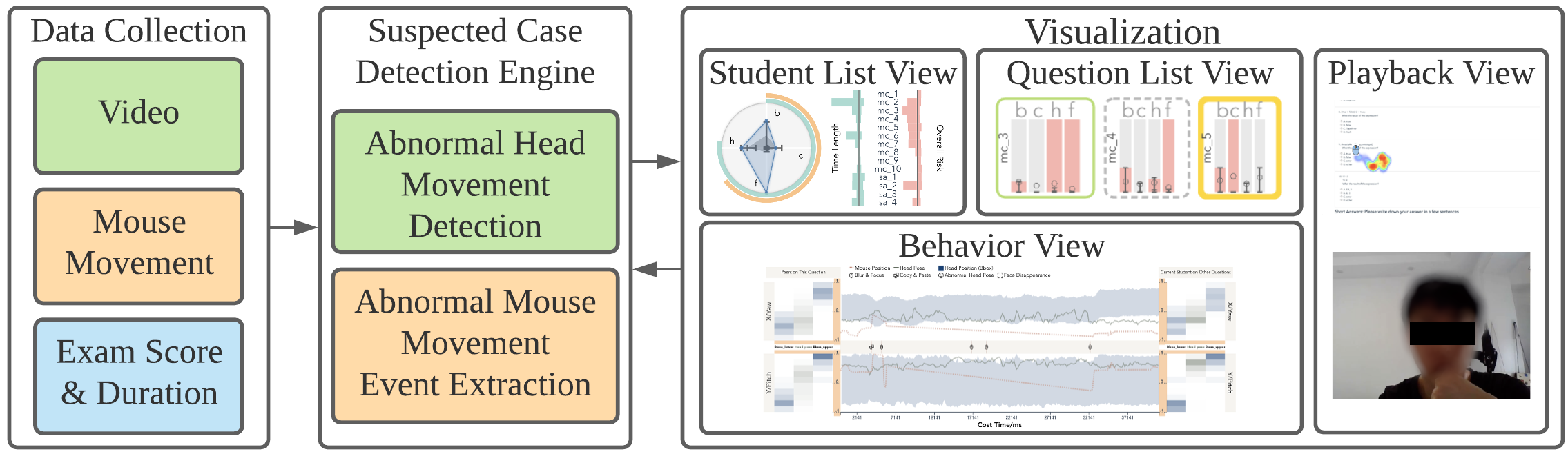}
    \vspace{-1em}
    \caption{
    Our method consists of three modules: data collection, suspected case detection engine and visualization.
    In the data collection module, each student's video, mouse movement data, exam score and duration are collected. These data are fed into the suspected case detection engine to extract abnormal head and mouse movements. The visualization module enables a convenient and efficient analysis 
    of students' online exam behaviors.
    }
    \Description[This figure shows the system overview.]{Our method consists of three modules: data collection, suspected case detection engine and visualization.
    In the data collection module, each student's video, mouse movement data, exam score and duration are collected. These data are fed into the suspected case detection engine to extract abnormal head and mouse movements. The visualization module enables a convenient and efficient analysis of students' online exam behaviors. }
    
    \label{fig:system_overview}
    \vspace{-1em}
\end{figure}

As shown in Figure~\ref{fig:system_overview}, the proposed approach consists of three major modules: data collection, suspected case detection engine and visualization, where the latter two modules are the core parts of our approach.
For the data collection, we mainly collected exam videos of individual students, mouse movement data and other related information like exam score and exam duration, which has been introduced in Section~\ref{section:data_collection}.
%
%
Then, in the suspected case detection engine, we conduct face detection and head pose estimation on videos to detect abnormal head movements.
We also define two types of abnormal mouse movements and further identify them from the mouse movement data.
Finally, we visualize the abnormal head and mouse movements and other related information with different levels of details.
There are four main views in our visualization module: \textit{Student List View} provides proctors with a quick overview of students at a high risk of cheating during the whole online exam~(Figure~\ref{fig:system}(a)); \textit{Question List View} facilitates the selection of high-risk periods of each student~(Figure~\ref{fig:system}(b)); \textit{Behavior View} presents the students' detailed behaviors of head and mouse movements~(Figure~\ref{fig:system}(c)); \textit{Playback View} makes proctors able to conduct a final confirmation on students' videos and mouse movements~(Figure~\ref{fig:system}(d)).

\subsection{Suspected Case Detection Engine}
\label{section:suspicious_engine}
We design a rule-based suspected case detection engine to identify suspected cases from both video and mouse movement data, and further estimate the risk of cheating. Specifically, we propose two rules: head poses which vary greatly from others and face disappearances in the video are abnormal head movements; copy, paste, blur and focus are abnormal mouse movements, since they can indicate the existence of cheating, as mentioned in Section~\ref{sec:collected_data}.



\begin{figure}[hbt!]
    \centering
    \includegraphics[width=0.5\linewidth]{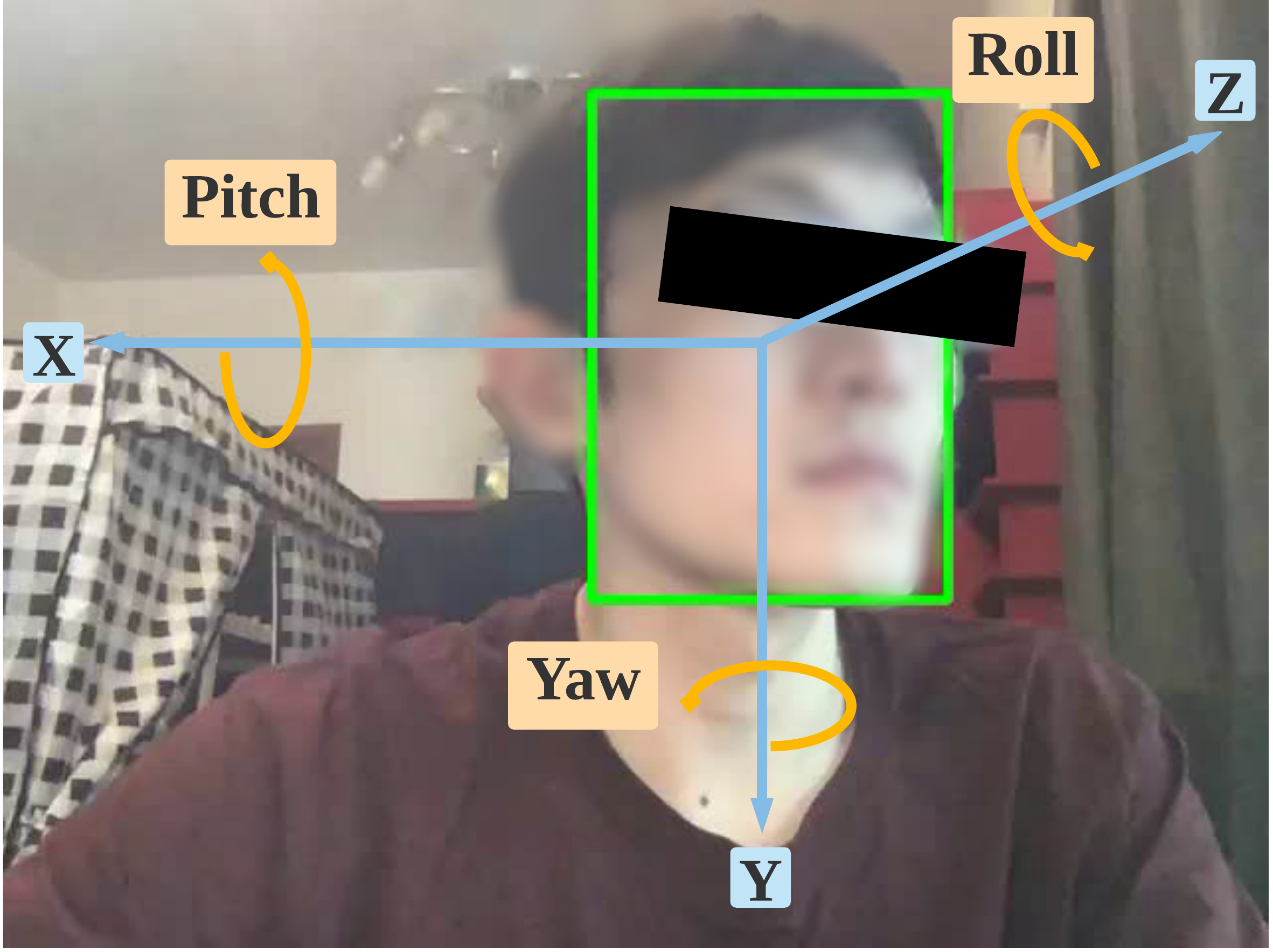}
    \caption{This figure illustrates the angles of the head pose and the head position. Pitch, yaw and roll describe the head rotation about X-axis, Y-axis, Z-axis, respectively. X-axis points to the right of the student. Y-axis points to the floor. Z-axis points to the computer screen. The green solid box is the bounding box representing the head position.}
    \label{fig:head_pose}
    \Description[This figure illustrates the angles of the head pose and the head position.]{Pitch, yaw and roll describe the head rotation about X-axis, Y-axis, Z-axis respectively. X-axis points to the right of the student. Y-axis points to the floor. Z-axis points to the computer screen. The green box is the bounding box representing the head position.}
    \vspace{-1em}
\end{figure}

\textbf{Abnormal Head Movement Detection.}
We characterize head movements from two perspectives: head poses and head positions.
Head poses indicate where a student is looking at during the online exam. 
Meanwhile, head positions represent how a student moves his/her head to use different devices or materials.
We use the position of a student's face in the collected video to delineate the corresponding head position, which is labeled as a rectangular bounding box that appropriately encloses the student's face, as indicated by the green box in Figure~\ref{fig:head_pose}. 
The size of a bounding box can help proctors check any change in distance between a student's face and the screen, which has been considered to be a useful indicator of cheating in prior research~\cite{chuang2017timedelay}. 
Also, the head position information can benefit the detection of cases when a student is not looking at the exam web page. 

In our approach, we focus on detecting two types of abnormal head movements: \textit{face disappearance} and \textit{abnormal head pose}, which are extracted by considering head positions and head poses, respectively.
\textit{Face disappearance} represents that the head position is unavailable during a period of time, since no face is detected. 
\haotian{This can happen under two circumstances where faces are not captured by webcams or are captured by webcams but not detected by the algorithm.
The situation where faces are not captured by webcams often result from the student leaving the room or covering the camera.
While an existing face being undetected by the algorithm usually happens when the student's face is partially covered by some objects, for example, a cup, or the student bows his/her head deeply.
According to our experts, P1 and P4, both cases often indicate a high probability of cheating. To verify if a case of face disappearance indicates cheating, proctors can check other information such as raw videos and mouse movement data. For example, in Scenario 1 of Section~\ref{usage_scenario}, we illustrate that drinking water is mistakenly judged as a face disappearance case and the raw video in our Playback View can help teachers verify it.}
An \textit{abnormal head pose} indicates that the head pose of the student varies greatly from a normal one.
For example, a student raises or bows his/her head, or turns his/her head away from the screen.

For abnormal head movement detection, we first extract head positions by conducting face detection in the video. 
Since there will be a large number of frames in each video,
we surveyed prior studies~\cite{zeng2020emoco, wu2020multiTED, zeng2019emotioncue} and follow Zeng~\textit{et al.} ~\cite{zeng2019emotioncue} by sampling video frames to accelerate video processing. Specifically, we process one video frame for every five frames.
Then, we use the pre-trained Faster R-CNN model~\cite{ruiz2017dockerface} to detect faces in the video. 
The extracted head position in Frame $i$ is labeled as a vector $[x_i^{min}, y_i^{min}, x_i^{max}, y_i^{max}]$,
where $(x_i^{min}, y_i^{min})$ is the coordinate of the upper-left corner and $(x_i^{max}, y_i^{max})$ is the coordinate of the lower-right corner. 
Also, the model outputs a probability score of correct detection for each detected head position to show if the detection is reliable.
To keep our results reliable, only the head positions 
with a probability
larger than $0.95$ are kept.
If Student $s$'s face is not detected in Frame $i$ in his/her video after sampling, then that frame is labeled as ``face disappearance''.
Student $s$'s total number of face disappearances is recorded as $n_{f}^s$.

Furthermore, we apply the state-of-the-art head pose estimation model~\cite{ruiz2018hopenet} to extract the head pose in each frame of the exam video of every student.
The extracted head pose at Frame $i$ of Student $s$'s video is a three dimension vector $[pitch_i^s, yaw_i^s, roll_i^s]$, which represents the head rotation about 3 axes (Figure~\ref{fig:head_pose}). 
Among all 3 angles, pitch and yaw angles are crucial to suspected case detection, since the pitch angle indicates where the student looks vertically and the yaw angle indicates where the student looks horizontally. However, the roll angle is not meaningful, since the head rotation along the Z-axis cannot affect much where the student is looking. 
Thus, it is not considered in our approach.
Since different students may have different exam settings, which lead to different ranges of head positions and head poses, we normalize each student's head positions and poses at each frame to $(-1, 1)$ with min-max normalization. 
In the rest of this paper, all the head poses $[pitch_i^s, yaw_i^s, roll_i^s]$ and head positions $[x_i^{min}, y_i^{min}, x_i^{max}, y_i^{max}]$ are the normalized ones.

Since students may place their webcam at various places or angles, a standard head pose is not available. 
Thus, we propose to use z-score\footnote{\href{https://en.wikipedia.org/wiki/Standard\_score}{https://en.wikipedia.org/wiki/Standard\_score}} to detect abnormal head poses based on the assumption that a student's head pose distribution should follow a normal distribution. We first calculate the average head pose of Student $s$ in his/her video, $[\overline{pitch^s}, \overline{yaw^s}]$, as well as the standard deviation of his/her head poses, $[\sigma_{pitch^s}, \sigma_{yaw^s}]$. 
Then a vector of z-scores of the head pose at Frame $i$ is calculated as 
\begin{equation}
    [\frac{pitch_i^s-\overline{pitch^s}}{\sigma_{pitch^s}}, \frac{yaw_i^s-\overline{yaw^s}}{\sigma_{yaw^s}}].
\end{equation}
If the absolute value of any z-score in the vector is larger than a threshold, the corresponding head pose is labeled as abnormal.
The default threshold is set to 3 by following the widely used Three-sigma rule\footnote{\href{https://encyclopediaofmath.org/wiki/Three-sigma_rule}{https://encyclopediaofmath.org/wiki/Three-sigma\_rule}} and proctors can interactively adjust the threshold through our visualization module. The total number of abnormal head poses of Student $s$ is recorded as $n_{h}^s$.

\textbf{Abnormal Mouse Movement Identification.}
We also detect a few representative suspected mouse movements, including ``copy'', ``paste'', ``blur'' and ``focus'', which can indicate cheating behaviors.
For example,
a ``copy'' can reveal that the student copies question contents and searches it for answers \haotian{if the ``copy'' is recorded before leaving the exam web page.} ``Paste'' sometimes signifies that a student pastes the answer from other sources, such as another website or electronic lecture notes.
\haotian{Specifically, if a ``paste'' is recorded after getting back to the exam web page, the student is most likely cheating}. 
``Blur'' and ``focus'' can indicate that a student leaves the current exam web page and gets back later. Since ``copy'' and ``paste'' are highly related to each other, we count the total number of ``copy''s and ``paste''s of Student $s$ as $n_{c}^s$. For the same reason, ``blur'' and ``focus'' are also counted together as $n_{b}^s$. Besides, following the same method of the normalization used on head poses and bounding boxes, we also normalize each student's mouse positions to $(-1, 1)$.

\textbf{Overall Risk Estimation.}
With all the suspected cases,
we further estimate students' risk levels on each question. First, all the occurrence numbers of suspected cases on Question $q$, $[n_{f}^{sq}, n_{h}^{sq}, n_{c}^{sq}, n_{b}^{sq}]$, are normalized to $(0, 1)$ with min-max normalization. 
For each type of suspected cheating behaviors on Question $q$, the minimum number of instances of the type is transformed to 0, while the maximum number of instances is transformed to 1.
In the rest of this paper, all the values $[n_{f}^{sq}, n_{h}^{sq}, n_{c}^{sq}, n_{b}^{sq}]$ 
are the normalized ones.
Finally, we summarized the overall risk of Student $s$ on Question $q$ as follows:
\begin{equation}
\label{equ:overall_risk}
    risk^{sq} = \sum_{t\in \{f, h, c, b\}} w_{t}\times n_{t}^{sq},
\end{equation}
where $w_t$ is a customized weight of each type of suspected cheating behaviors. The weights of each type are all set to 1 by default and our approach also enables proctors to interactively adjust them.

\subsection{Visual Design}

As mentioned above, we also proposed straightforward and intuitive visual designs to help proctors explore and analyze student cheating behaviors based on student exam videos and mouse movements. 

\begin{figure*}[hbt!]
    \centering
    \includegraphics[width=\linewidth]{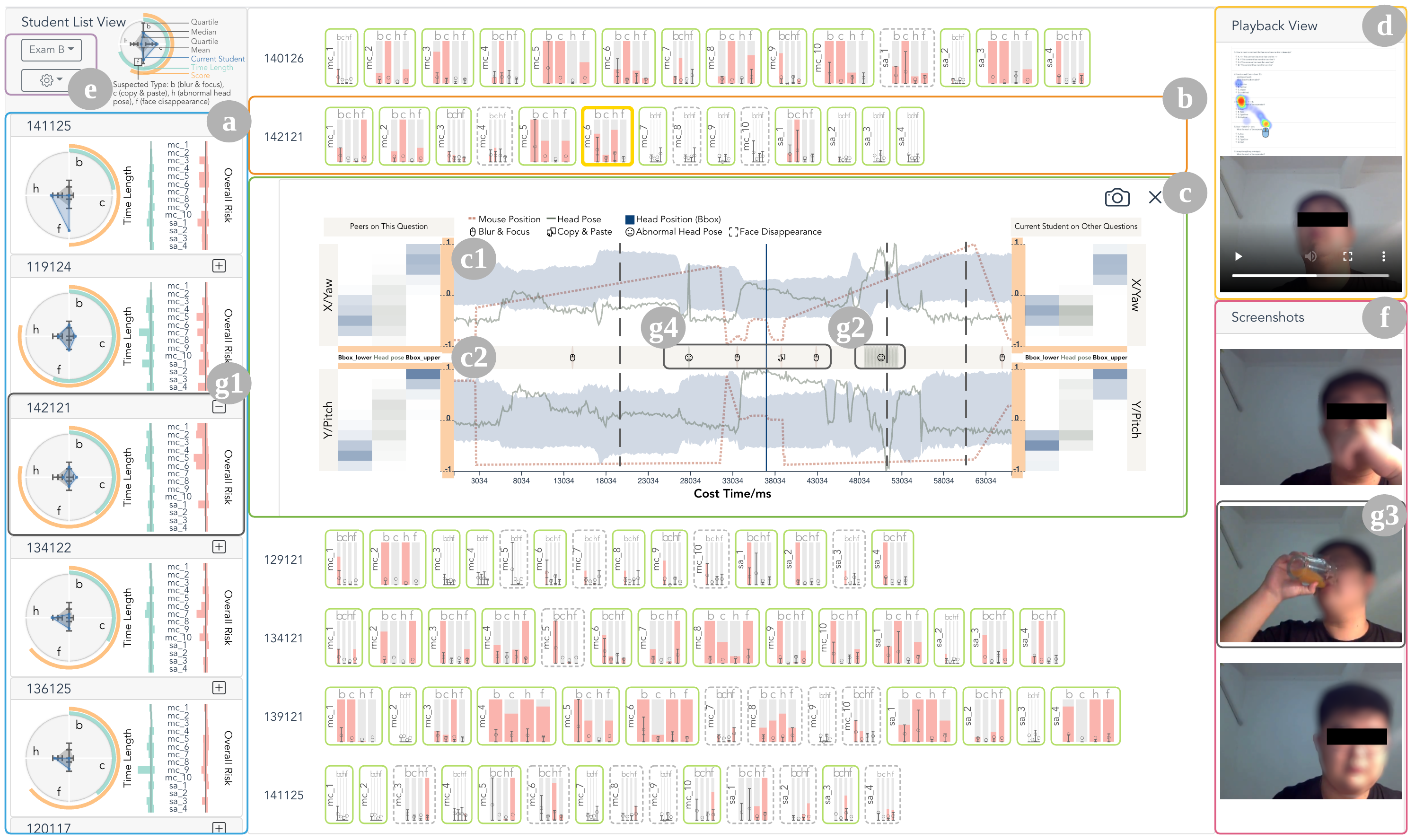}
    \caption{Our approach provides novel visualizations for proctors to identify cheating cases. (a) \textit{Student List View} is an overview of all students' risk of cheating in an online exam. (b) \textit{Question List View} shows the risk level of all questions finished by a student. (c) \textit{Behavior View} presents a student's detailed head and mouse movements while answering a question. (c1) and (c2) are the upper detailed behavior chart and the suspected case chart, respectively. (d) \textit{Playback View} enables proctors to check raw videos and animated visualization of mouse movements on the exam web page. (e) The control panel can be used to select online exams and adjust several parameters by proctors. (f) provides a function to save screenshots of the raw video. Screenshots in (f) are taken at time points indicated by vertical black dashed lines in (c). (g1)-(g4) illustrate fast location and convenient verification of cheating behaviors in \textit{Usage Scenario 1}.} 
    \label{fig:system}
    \Description[This figure shows the novel visual interface for proctors to identify cheating cases.]{(a) Student List View on the left is an overview of all students' risk of cheating in an online exam. (b) Question List View in the middle shows the risk level of all questions finished by a student. (c) Behavior View  presents a student's detailed head and mouse movements during answering a question. It is shown under each student's Question List View and can be collapsed.  (c1) and (c2) are the detailed behavior chart and the suspected case chart, respectively. (d) Playback View in the top right corner enables proctors to check raw videos and animated visualization of mouse movements on the web page. (e) The control panel on the top of Student List View can be used to select online exams and adjust several parameters by proctors. (f) provides a function to save screenshots of the raw video. Screenshots in (f) are taken at time points indicated by vertical black dashed lines in (c).  (g1)-(g4) illustrate fast location and convenient verification of cheating behaviors in Usage Scenario 1.}
\end{figure*}

\textbf{Student List View.} 
Student List View (Figure~\ref{fig:system}(a)) is designed to provide proctors with an overview of the risk levels of all students who have participated in an online exam~(\textbf{R1}). 

\begin{figure}[hbt!]
    \centering
    \vspace{-1em}
    \includegraphics[width=0.9\linewidth]{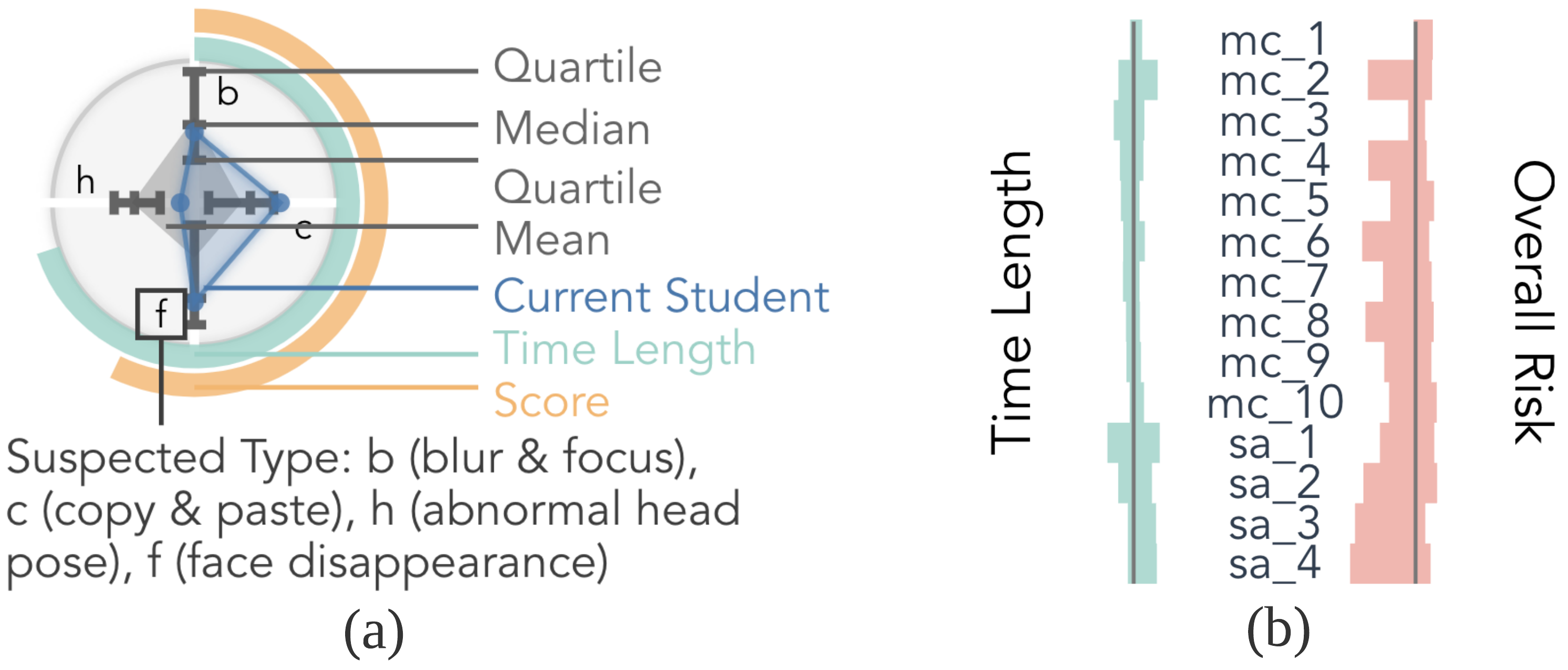}
    \vspace{-1em}
    \caption{Overview of a students' risk of cheating is shown by two main components: (a) a glyph showing overall risk of all suspected types; (b) two diverging bar charts showing the comparison of current student's and all students' overall risk and time spent on each question.}
    \label{fig:student_view}
    \Description[This figure shows two main components of the overview of a students' risk of cheating.]{The two main components are: (a) a glyph showing overall risk of all suspected types composed of two inner radar charts and two outer radial bar charts; (b) two diverging bar charts showing the comparison of current student's and all students' overall risk and time spent on each question.}
        \vspace{-1em}
\end{figure}

Each row of Student List View visualizes the risk of a student,
which is composed of two main parts, a glyph and two diverging bar charts. The glyph (Figure~\ref{fig:student_view}(a)) shows the overall risk of suspected types and the diverging bar charts (Figure~\ref{fig:student_view}(b)) display the difference of cheating risks and time costs between the current student and all the students.
The glyph has two outer radial bar charts and two radar charts.
\haotian{The blue radar chart with circles on vertices encodes the current student's normalized risk level of each suspected type in a spatial position.
The grey radar chart without marks on vertices indicates the average normalized risk level of all students in this online exam.
The green outer radial bar chart of a smaller radius shows the percentage of time used to finish the online exam, while 
the orange radial bar chart of the larger radius shows the percentage of scores.}
This design follows the effectiveness principle described by Munzner~\cite{munzner2014visualization} by encoding the most important information, cheating risk, using a strong visual channel (i.e., spatial position) and encoding the less important information, time used and scores, with a weaker channel (i.e., angle).
Also, we 
apply a boxplot-like design and  
visualize the 1st, 2nd and 3rd quartiles\footnote{\href{https://en.wikipedia.org/wiki/Quartile}{https://en.wikipedia.org/wiki/Quartile}} of the normalized risk levels on each axis to provide
proctors with more detailed analysis and enable easy comparison of current student and other peer students~(\textbf{R4}). 
The range of each axis is $(0, 1)$ from the center to the edge.
An alternative design for plotting these statistical metrics, which is considered during our design process, is to draw all of them using radar charts. However, 
it will lead to severe occlusion if five radar charts are drawn within the glyph.
Thus, we finally adopt the boxplot-like design to show these metrics.

Two diverging bar charts show the time spent on each question in green on the left and each question's overall estimated risk in red on the right. The risk of each question is calculated as Equation~\ref{equ:overall_risk}. The right side of both bar charts shows the average normalized value of all students, and their left side presents the normalized value of current students. 
The length of each bar encodes the normalized value.
The design of diverging bar charts also aims to achieve
a convenient comparison of a student and others~(\textbf{R4}). 

Rich interactions are supported in Student List View. 
First, a tooltip will display the normalized risk level when hovering on the radar chart.
Second, the control panel (Figure~\ref{fig:system}(e)) enables proctors to adjust a set of configurable parameters for risk estimation, such as the threshold of abnormal head poses and the weights of different suspected cheating types, as mentioned in Section~\ref{section:suspicious_engine}. The method of sorting and the selection of online exams can also be changed in the control panel.
Third, proctors can click on the ``plus'' icon on the upper-left corner of each row to show the current student's Question List View. 


\textbf{Question List View.}
Question List View~(Figure~\ref{fig:system}(b)) is designed to help proctors quickly locate high risk questions, where a student may have cheated,
for further investigation~(\textbf{R2}).

In this view, each question is represented by a block. \haotian{The color and border style encode whether the student correctly answers a particular question, where green solid rectangles represent that the corresponding questions are correctly answered and gray dashed rectangles indicate incorrect answers.}
The width of the block encodes the estimated risk level of one question, which makes high risk questions easier to be noticed by proctors. 
Within each block, there is a bar chart showing all normalized risk levels of the four suspected cheating types, ``blur and focus'' (b), ``copy and paste'' (p), ``abnormal head pose'' (h) and ``face disappearance'' (f).
To support a convenient comparison with all the other students, we 
apply a boxplot-like design to indicate the 1st, 2nd and 3rd quartiles of the normalized risk levels~(\textbf{R4}). 
The circle on each bar encodes the average cheating risk level among all students on the same question.
Proctors can click a question block of the student and expand the Behavior View of that question for detailed inspection. 
\haotian{If a question is selected, it will 
be highlighted by a thicker gold solid border
for easy identification. }

During our design process, we initially considered
using a heatmap to visualize the cheating risk of questions of each student in the Student List View. Each block in the heatmap represents a question in the online exam and the opacity of each block denotes the risk of each question. 
However, this design suffers from some disadvantages.
First, detailed risk distribution on different types of suspected cases can hardly be represented in a block in heatmaps. Second, due to the limited screen space, scalability is also a concern of such a design, especially when there is a large number of questions in an online exam. Thus, inspired by EgoSlider~\cite{wu2016egoslider}, we finally decided to use diverging bar charts to show the overall risk of each question in the Student List View and further design Question List Views to present the detailed risk of questions.

\textbf{Behavior View.}
\label{sec:behavior_view}
Behavior View (Figure~\ref{fig:system}(c)) provides proctors with a detailed understanding of 
how the student's head and mouse move during his/her problem-solving process,
which
enables the
fast location of suspected cases and further inspection~(\textbf{R3}). 
This view consists of three types of charts: two \textit{detailed behavior charts} in the middle, a \textit{suspected case chart} between two detailed behavior charts and periphery heatmaps on the left and right sides. The detailed behavior charts are plotted to show the detailed head and mouse movements, while the periphery heatmaps~\cite{morrow2019periphery} are used for comparing the behavior shown in detailed behavior charts across students and questions. 

In Figure~\ref{fig:system}(c), the upper 
detailed behavior chart~(Figure~\ref{fig:system}(c1)) presents the mouse positions and ranges of bounding boxes along the X-axis,
while the lower
detailed behavior chart~(Figure~\ref{fig:system}(c2)) shows the corresponding information along the Y-axis.
The yaw angles of head poses are shown in the upper detailed behavior chart and the pitch angles are shown in the lower detailed behavior chart. 
\haotian{In both detailed behavior charts, the brown dashed line shows the normalized mouse positions on the screen, while the dark green solid line encodes the normalized angles of head poses.} 
The area in light blue represents the range of bounding boxes of head positions on one axis, which means the upper bound of the area is $x_{max}$ or $y_{max}$, while the lower bound represents $x_{min}$ or $y_{min}$.
However, as pointed out by Blascheck~\textit{et al.}~\cite{Blascheck2017eyetracking_vis}, encoding movement data by axes individually has a drawback that extra mental effort is needed to understand movements. 
To mitigate this problem, we also provide animated visualizations of mouse movements and raw videos in Playback View to help proctors understand how a student has behaved during the online exam (Figure~\ref{fig:system}(d)).
The suspected case chart~(Figure~\ref{fig:system}(c2)) between two detailed behavior charts shows the positions of all suspected cases. 
We plot a bar for each suspected case detected by our suspected case detection engine 
and the glyph on the bar denotes the type of suspected cheating cases. 

\begin{figure}[hbt!]
    \centering
    \includegraphics[width=0.7\linewidth]{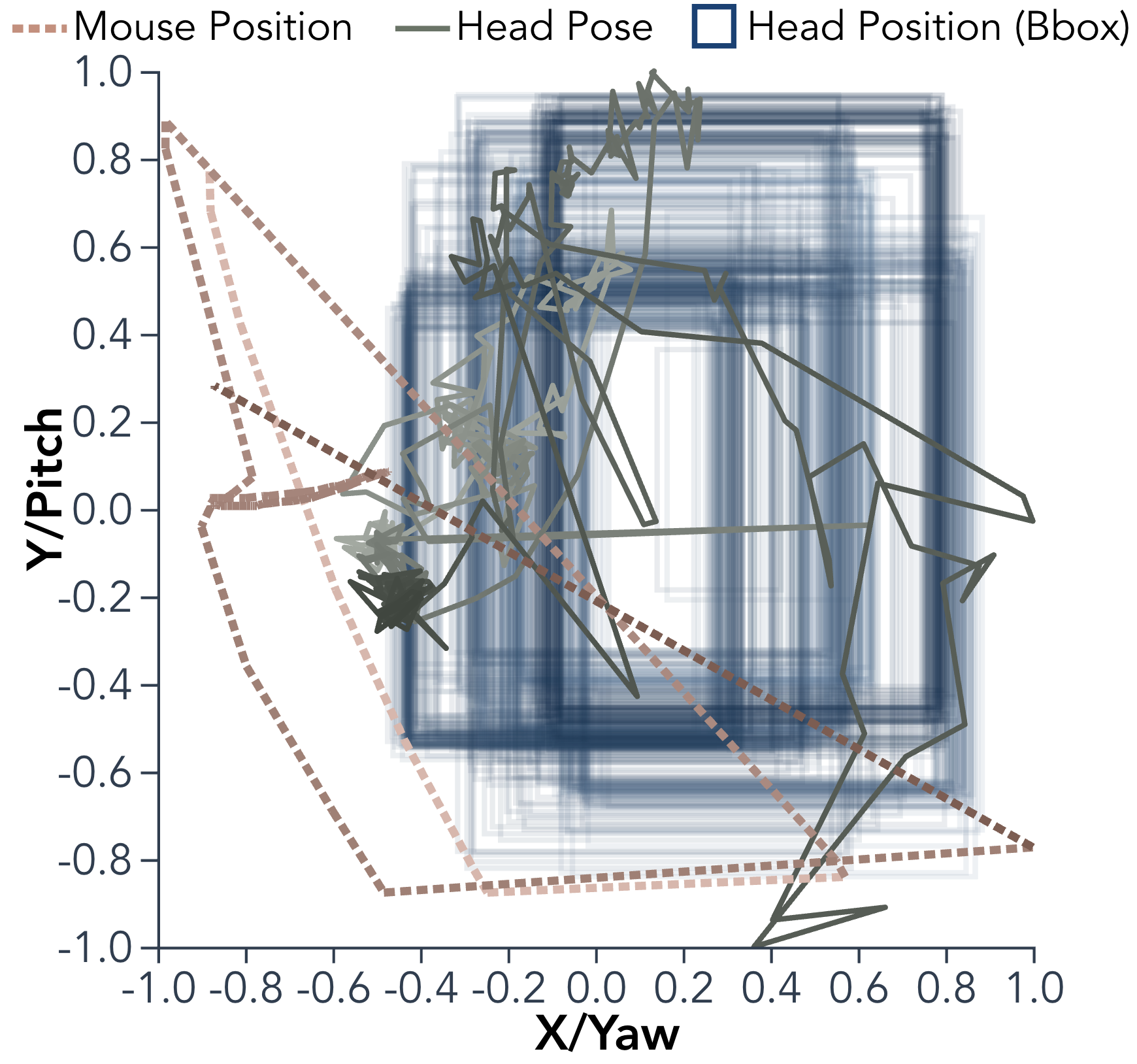}
        \vspace{-1em}
    \caption{An alternative design of detailed behavior charts. It shows the same head and mouse movements as these in Figure~\ref{fig:system}(c). \haotian{The brown dashed line and dark green solid line represent mouse movements and head poses respectively. Blue solid boxes encode head positions.}}
    \Description[This figure shows an alternative design of detailed behavior charts.]{It shows the same head and mouse movements as these in Figure 3(c). The brown dashed line and dark green solid line represent mouse movements and head poses respectively. Blue boxes encode head positions. This design leads to severe visual clutter when the amount of data is huge.
   
    }
    \label{fig:alternative_design}
     \vspace{-1em}
\end{figure}

Before adopting the current design to present head and mouse movement data, we also considered a possible alternative design.
Inspired by the scanpath visualizations of eye tracking data~\cite{Blascheck2017eyetracking_vis}, we propose a visual design to show the original X-axis (horizontal axis) and Y-axis (vertical axis) coordinates of head positions and mouse positions, as shown in Figure~\ref{fig:alternative_design}. 
Also, the yaw angles of head poses are shown together with the X-axis, since it reflects where the student is looking at on the X-axis. For the same reason, the pitch angles are shown together with the Y-axis. 
In this design, the trajectory of mouse movements and the change of head poses are shown as lines. The head positions are represented as bounding boxes.
The opacity of lines and box borders encodes the temporal information, where a high opacity indicates that the head or mouse movement occurs at the latter stage of the whole problem-solving process for this question.
However, this design leads to severe visual clutter and occlusion when the amount of data is huge. 
The visual clutter further makes it hard to learn the sequence of movements.
To reduce the visual clutter and better encode the temporal information for easy location of suspected cases, we finally
decide to employ our current design, which breaks down
the spatial movement information into two  dimensions and encodes them individually.

On the left and right sides of the detailed behavior charts, four periphery heatmaps which have the same Y-axes as those of detailed behavior charts are designed to display other students' behaviors on the current question and the current student's behaviors on other questions, respectively.
In each heatmap, there are three columns which represent the frequency distributions of the lower bounds of head positions, head poses and the upper bounds of head positions from left to right.
The color of each column is the same as the corresponding line chart in the detailed behavior chart. 
Its opacity encodes the frequency of head poses or head positions that fall into a specific interval (e.g., the frequency of head poses with a value between 0 and 0.1).
The heatmaps are designed to facilitate comparison of student behaviors across students and questions~(\textbf{R4}).
Such comparisons are necessary to a reliable cheating behavior analysis, since they enable proctors to consider students' habits and the specific questions they are working on, which are two important factors affecting student behaviors.

\textbf{Playback View.}
Playback View (Figure~\ref{fig:system}(d))
provides
proctors with a choice to review the suspected cases and further confirm whether a suspected case is real cheating, especially for some ambiguous cases~(\textbf{R5}).
This is important, as the underlying suspected case detection algorithm 
often cannot achieve a 100\% accuracy and some normal behaviors can be misclassified as cheating, for example, drinking water.
Also, it serves as a complement to the detailed behavior charts in Behavior View.
The view contains two parts, an animated mouse movement visualization at the top and a raw video player at the bottom. 
The animated mouse movement visualization uses a heatmap to show the number of times that the mouse stays in an area.
In the color scale of our heatmap, blue denotes few visits and red denotes frequent visits.
\haotian{Also, The opacity of an area with fewer visits is lower.}
In Playback View, the raw video player and the mouse movement animated visualization are linked together to play synchronously.
Additionally, they can be controlled by clicking on the detailed behavior charts to skip to a certain time point and start to play.
The vertical blue solid line in Figure~\ref{fig:system}(c) denotes the current time point of playing the raw video and the animated visualization.
The raw video player supports multiple interactions, including play/pause, skip and play in full-screen.
Furthermore, proctors can click the ``camera'' button at the top right corner of Behavior View (Figure~\ref{fig:system}(c)) to take a screenshot of the current video and list some screenshots in the area below Playback View for an easy review of videos, as shown in Figure~\ref{fig:system}(f).


\section{Evaluation}
We extensively evaluate our approach through three usage scenarios, a user study and expert interviews.

\vspace{-1em}
\subsection{Usage Scenario}
\label{usage_scenario}
In this section, we describe three usage scenarios to demonstrate the usefulness and effectiveness of our visual analytics system in facilitating the proctoring of online exams. 

\textbf{Scenario 1: Fast Location and Convenient Verification of Cheating Behaviors} 

In this scenario, we report a whole workflow to find a cheating case by observing mouse movements in a convenient and reliable manner. 
First, we select ``Exam B'' in the control panel and sort students according to their level of risk. 
Then, we browse the Student List View to find students of high risk. 
Among these students, the student whose ID is \textit{142121} is found that his risks of several types are higher than median values in Figure~\ref{fig:system}(g1)~(\textbf{R4}). 
To further confirm if he really cheated in the online exam, we expand his Question List View and locate the two most suspected questions, $mc\_5$ and $mc\_6$ by observing the widths of blocks and comparing the overall risk level with other students in the Student List View~(\textbf{R2, R4}).
In the block of $mc\_6$, we find there are a large number of suspected cases of abnormal head pose and ``blur and focus''. Then, we click on the block to further investigate the detailed behavior while answering that question in Behavior View~(\textbf{R3}).
In Behavior View on this question, we first notice that there are multiple abnormal head poses near the end of the question-answering process~(Figure~\ref{fig:system}(g2)) in the suspected case chart. 
The line charts of head poses are further compared with the heatmaps on both sides to confirm that these abnormal head poses are not led by the question or the habits of \textit{142121}~(\textbf{R4}). 
Then we click on the detailed behavior charts at the beginning of those abnormal head poses to check the video in Playback View~(\textbf{R5}). 
In the video, we find that actually, he had drinks~(Figure~\ref{fig:system}(g3)).
Then these abnormal head poses can be considered as normal behavior at low risk of cheating.
We also notice that there are several ``blur and focus''s and a ``copy and paste'' during his period of answering $mc\_6$ in ~(Figure~\ref{fig:system}(g4)). 
To further confirm all ``copy and paste'' cases are not conducted on the current page, we observe the line charts of mouse positions. These charts suggest that he arrived at the boundary of the web page and stayed for a while. Then we feel quite confident that he left the web page and copied and pasted some materials, which is considered to be a cheating case in our online exam setting~(\textbf{R3}).

This scenario demonstrates that our system can help proctors quickly locate and verify suspected cases. Also, it shows that mouse movement data provides a new perspective to find cheating cases.


\textbf{Scenario 2: Cheating Cases Identification Through Detailed Inspection of Head Movements}

\begin{figure}[hbt!]
    \centering
    \includegraphics[width=\linewidth]{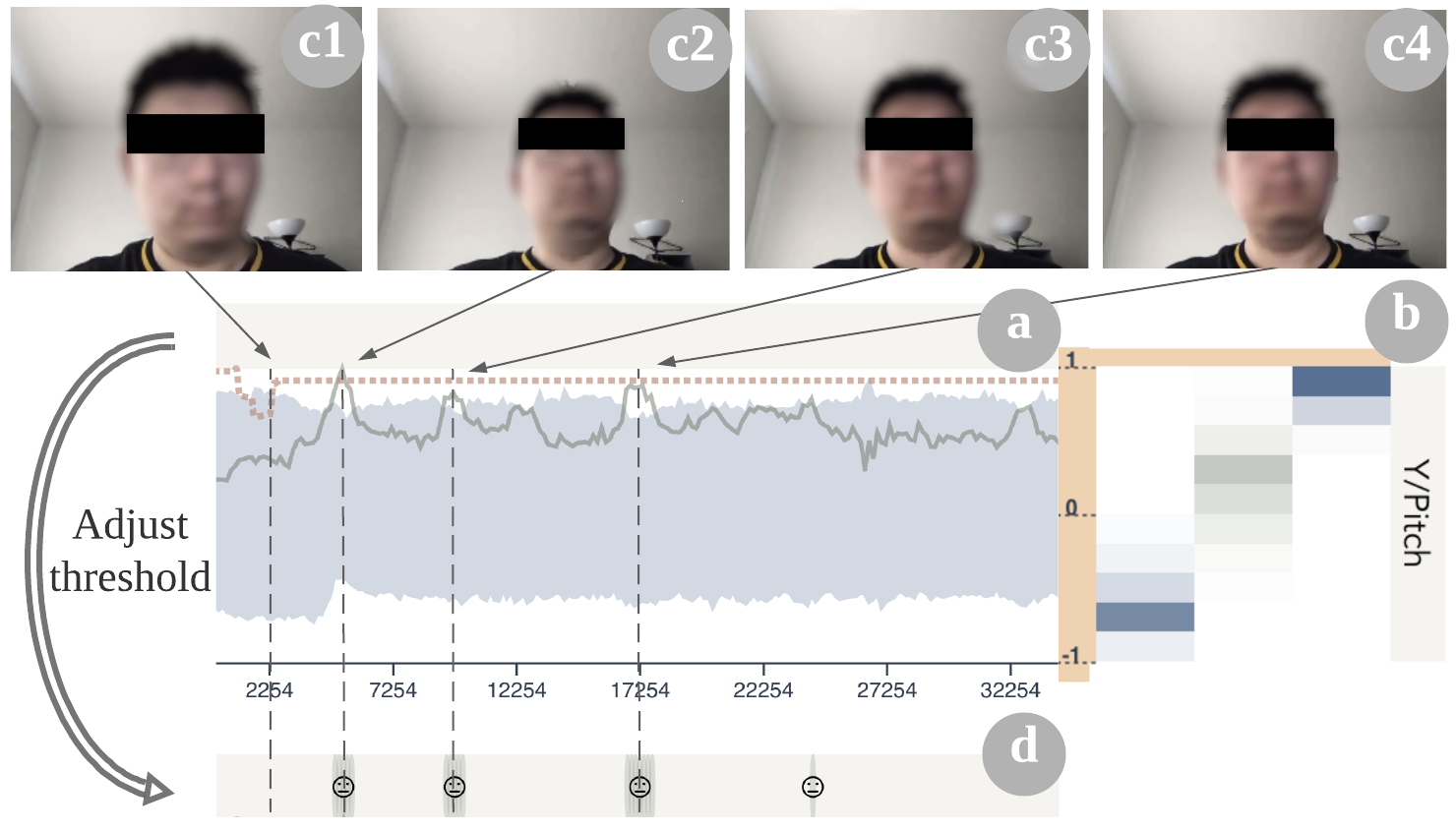}
    \vspace{-1em}
    \caption{This figure illustrates how to identify cheating cases through detailed inspection of head movements in \textit{Usage Scenario 2}. (a) shows part of the detailed behavior chart on Y-axis and pitch angles and the suspected case chart in Behavior View (threshold of abnormal head poses = 3). (b) shows the periphery heatmap of the current student on other questions.   (c1)-(c4) are the screenshots while answering the question and vertical black dashed lines indicate time points of screenshots. (d) shows the same part of the suspected case chart after threshold adjustment~(threshold of abnormal head poses = 2).}
    \Description[This figure shows illustrates how to identify cheating cases through detailed inspection of head movements in Usage Scenario 2.]{(a) shows part of the detailed behavior chart on Y-axis and pitch angles and the suspected case chart in Behavior View (threshold of abnormal head poses = 3). (b) shows the periphery heatmap of the current student on other questions.   (c1)-(c4) are the screenshots while answering the question and vertical black dashed lines indicate time points of screenshots. (d) shows the same part of the suspected case chart after threshold adjustment (threshold of abnormal head poses = 2).}
    \label{fig:case_study_4}
    \vspace{-1em}
\end{figure}

In this scenario, we report a cheating case which is found through an in-depth inspection on head movements.
In the student's Behavior View (Figure~\ref{fig:case_study_4}(a)), we notice there is no suspected case detected by our detection engine during that period. Thus we may need to observe the detailed behavior charts to further learn if he has any risk of cheating . 
First, a sudden change of head positions after his mouse stops moving is noticed. 
The bounding box becomes smaller, which means the distance between his face and the screen is larger.
Then we check his line charts of head poses and find that almost at the same time, his head is raised and afterwards the pitch angles of his head poses are frequently outside his normal range of pitch angles by comparing the detailed behavior chart with the heatmap~(\textbf{R4}). 
Then we further check his video and find that he raised his head and seems to look at something other than the laptop screen~(\textbf{R5}), as Figures~\ref{fig:case_study_4}(c2)-(c4) show. 
Thus, this case is thought to be a potential cheating case. After checking his reported cheating behaviors, he tried to use another computer behind the laptop to search for answers, which matches our observation.


Since the detection engine fails to find these abnormal head poses, we would like to investigate the setting of the threshold. We use the control panel in Figure~\ref{fig:system}(e) to lower our threshold from the default value (i.e., 3) to 2, which means head poses of smaller variation than the default will also be considered as suspected cases and the proctoring will be more strict.
The suspected case chart after adjustment is shown in Figure~\ref{fig:case_study_4}(d). We can see some suspected cases appear in the suspected case chart. According to our observation, the positions of glyphs match the moments the student raised his head and looked somewhere other than the screen. Thus, we may consider that an inappropriate setting of the threshold on this student led to no suspected case being detected. However, since different online exams may have different requirements of the proctoring, it is hard to define a unified threshold of abnormal head poses. We leave this as an option for proctors to provide them with sufficient flexibility to detect suspected behaviors. Also, our suspected case detection only aims to provide references to proctors instead of directly making decisions about whether the student cheats in the online exam. Proctors need to further observe the detailed behaviors for final decisions.

\textbf{Scenario 3: Cheating Case Identification through the Inconsistency between Mouse and Head Movements} 

\begin{figure}[hbt!]
    \centering
    \includegraphics[width=\linewidth]{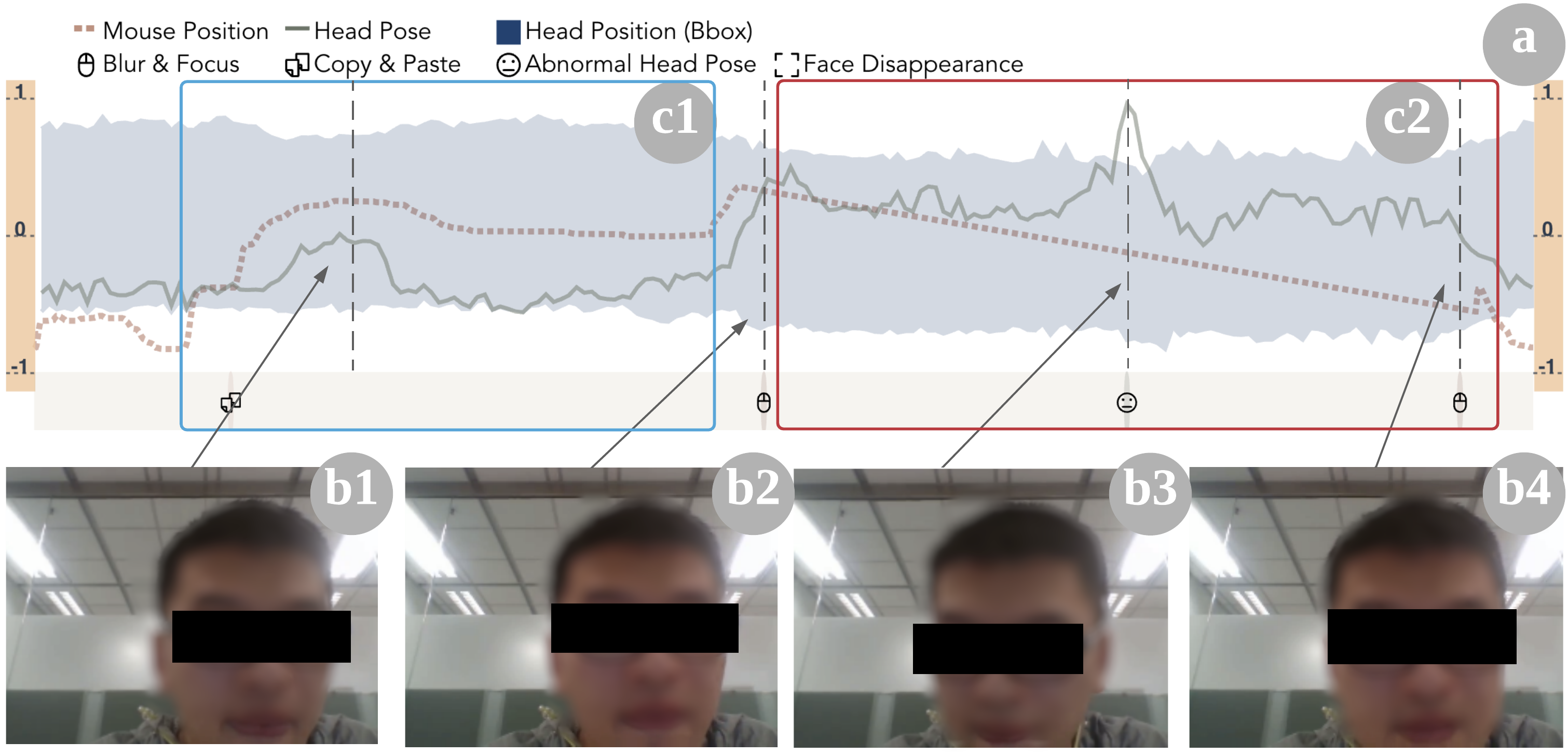}
    \caption{This figure shows the identification of a cheating case through the inconsistency between mouse and head movements in \textit{Usage Scenario 3}. (a) is the detailed behavior chart on X-axis and yaw angles in the Behavior View. (b1)-(b4) show the video screenshots while answering the question and vertical black dashed lines indicate time points of screenshots. (c1) shows consistency of head and mouse movements. (c2) shows the inconsistency of head and mouse movements.}
    \vspace{-1em}
    \label{fig:case_study_3}
    \Description[This figure shows the identification of a cheating case through the inconsistency between mouse and head movements in Usage Scenario 3.]{(a) is the detailed behavior chart on X-axis and yaw angles in the Behavior View. (b1)-(b4) show the video screenshots while answering the question and dash lines indicate time points of screenshots. (c1) shows consistency of head and mouse movements. (c2) shows the inconsistency of head and mouse movements.}
\end{figure}

In this scenario, a cheating case is identified through the analysis of the inconsistency between head and mouse movements~(\textbf{R3}) using our approach.
We first check the detailed behavior chart on X-axis and yaw angles in the Behavior View of a student (Figure~\ref{fig:case_study_3}(a)) and quickly notice that the student's head and mouse movements are consistent initially (Figure~\ref{fig:case_study_3}(c1)), but vary a lot in the latter stage
(Figure~\ref{fig:case_study_3}(c2)). 
From the consistency of his head and mouse movement behaviors in Figure~\ref{fig:case_study_3}(c1), we can learn that he kept looking at his cursor. Thus, the phenomenon in Figure~\ref{fig:case_study_3}(c2) can be abnormal.
It indicates that either the student does not look at the cursor or the student leaves the current web page, since
collecting mouse movement data on other web pages or applications is unavailable. 
We further investigate the suspected case chart and two suspected cases of ``blur and focus'' draw our attention. These two suspected cases happen at the beginning and the end of the period of the inconsistency, which confirms that the student left the exam web page.
By further considering the suspected case of ``copy and paste'' at the beginning of the period in Figure~\ref{fig:case_study_3}(c1), the whole cheating process can be inferred: the student copied the question content and searched it or ran it in an IDE (Integrated Development Environment).
%
By only viewing his video, his cheating behavior can hardly be detected, since there are almost no suspected head movements in the video, as Figures~\ref{fig:case_study_3}(b1)-(b4) show.

This scenario shows that our visualizations can enable easy identification of cheating behaviors by exploring the inconsistency between head and mouse movements,
which are not able to be revealed in videos.
Also, it demonstrates the effectiveness of introducing mouse movements to our proctoring system.

\subsection{User Study}

We also conducted a user study to further quantitatively assess the effectiveness of our approach for facilitating the proctoring of online exams.
Specifically, we evaluate the time cost and accuracy of finding cheating cases of our approach and compare it with the baseline approach (i.e., manually going through the exam videos).

\textbf{Datasets and Tasks.}
Two datasets in our user study were collected in our mock online exam, as described in Section~\ref{section:data_collection}. 
For each question set in our mock online exam, we picked 20 video clips of 6 students who cheated on that question set as a dataset. 
The total length of video clips in both datasets was around 15 minutes. 
In each dataset, according to reported cheating behaviors, we chose 10 video clips with cheating behaviors and another 10 video clips without cheating behaviors.
Besides, we selected 4 more video clips as our demo dataset. 
Each dataset contained instances of both cheating types defined in Section~\ref{section:cheating_types}.

\begin{figure}[hbt!]
    \centering
    \includegraphics[width=\linewidth]{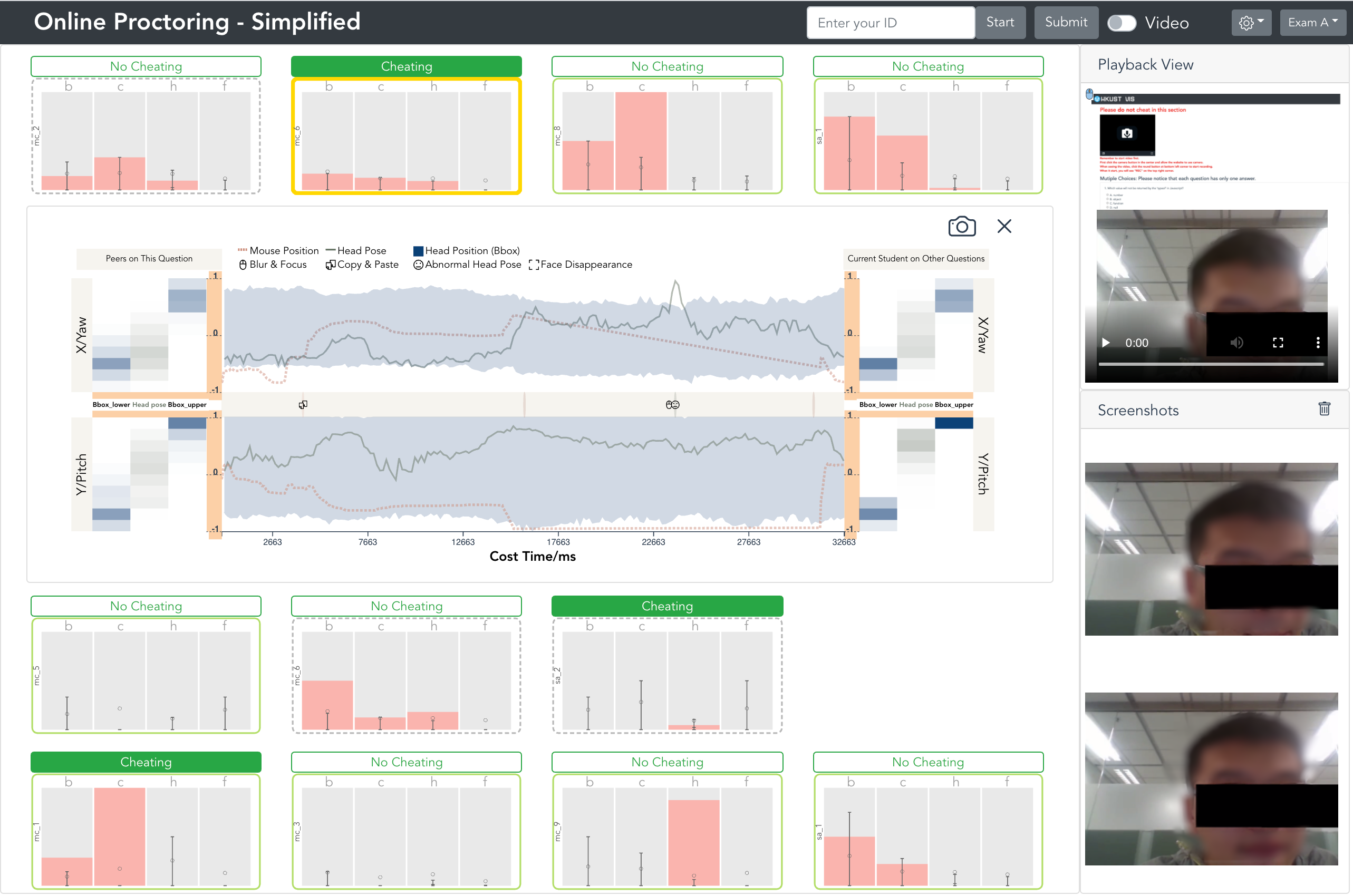}
    \caption{\haotian{The screenshot of the simplified system used in Task 1. Compared with the complete system, it removes Student List View and adds some tools to record results, e.g., the toolbar at the top and buttons above the questions. The toolbar is used to record the start and submission timestamps and the buttons allow participants to select cheating cases.}}
    \Description[This figure shows the screenshot of the simplified system used in Task 1.]{Compared with the complete system, it removes Student List View and adds some tools to record results, for example, the toolbar at the top and buttons above the questions. The toolbar is used to record the start and submission timestamps and the buttons allow participants to select cheating cases.}
    \label{fig:simplified_system}
\end{figure}

\haotian{
In our user study, each participant was asked to perform two tasks sequentially.
\textit{Task 1} is designed to compare the effectiveness and efficiency of our approach with the baseline approach for proctoring online exams, i.e., viewing the original exam videos of students. 
Each participant reviewed two datasets by viewing raw videos (i.e., the \textit{baseline} method) or using our system to label cheating cases, respectively.
In Task 1, to ensure a fair comparison, we showed each participant a simplified system with only Question List View, Behavior View and Playback View. 
The reason why we removed Student List View is that this view is designed to identify high risk students, while the baseline method does not provide such functionality.
A screenshot of the simplified system is shown in Figure~\ref{fig:simplified_system}.
The goal of \textit{Task 2} is to let participants try the complete workflow of our system and evaluate the usability and visual design of our entire system and individual views.
Thus, we asked participants to freely explore our complete system and finish a questionnaire at the end.}

\vspace{-1em}

\textbf{Participants.}
We recruited 16 postgraduate students (5 female, $age_{mean}=24.81$, $age_{sd}=2.10$) from various departments including Computer Science, Electronic Engineering, Economics and Environmental Science in a local university through word-of-mouth and social media. All participants are or have been teaching assistants~(TAs) in the university. Due to the current pandemic of COVID-19, the study was conducted in a blend mode. Some participants took our study face-to-face and others took through online meetings. After the completion of the study, each participant 
was compensated with US \$7.


\textbf{Procedure.}
The whole study lasted about 1 hour. \haotian{At the beginning, we briefly introduced the purpose of the user study and what data would be collected during the procedure. We asked for participants' permission to allow us to use the collected data for research purposes anonymously.}
Then we introduced the whole procedure and had the tutorial session.
In the tutorial session, we demonstrated the usage of our system, explained all suspected types and then
asked them to
try our simplified system with the demo dataset.
After the tutorial session, they conducted \haotian{Task 1} on different datasets by using the baseline method or using our simplified system. The order of reviewing methods and datasets was counterbalanced to eliminate the effect brought by the differences in the datasets. 
The time limit for each dataset was 10 minutes, but participants were allowed to submit their results early once they finished reviewing all the videos in the dataset and felt confident about their selections. 
The reason why our time limit was shorter than the total length of video clips is that we would like to mimic the real procedure of proctoring, in which proctors do not have time to review all videos and they sometimes skip some details to save time.
We recorded their time used as the time between clicking a ``Start'' button and submitting their results. 
\haotian{After submitting all reviewing results for Task 1, they were able to have a short break before starting Task 2. 
In Task 2,
we first gave a demo on the entire workflow and emphasized the design of Student List View since it was not shown in Task 1.
Then the participants spent around 20 minutes to freely explore our complete system with all collected data.
After the exploration, they were asked to finish a questionnaire to evaluate our system.
}
In our questionnaire, we adopted the bipolar survey design with negative statements at the left end of 5 scale points (1-5 with 1 as the most negative and 5 as the most positive) and positive statements at the right end. 
Also, there were two text questions for suggestions and comparisons with the baseline approach. All questions are listed in Table~\ref{table:questions}. 

\begin{table}[hbt!]
    \centering
    \footnotesize
    \caption{The first section of our questionnaire is designed to evaluate usability~($Q1$-$Q5$) and the visual design~($Q6$-$Q8$) of the whole visual analytics system.
    The second section of our questionnaire is designed to evaluate usefulness and usability of Student List View~($Q9$-$Q12$), Question List View~($Q13$-$Q14$) and Behavior View~($Q15$-$Q19$). The third section of our questionnaire is to ask about some personal opinions on our system~($Q20$-$Q21$). The original sentences without the words in brackets are the positive statements at the right end of the scale points, while the sentences with words in the brackets are the negative statements at the left end.}
    \Description[This table shows all questions in our questionnaire.]{The first section of our questionnaire is designed to evaluate usability (Q1-Q5) and the visual design (Q6-Q8) of the whole visual analytics system.
    The second section of our questionnaire is designed to evaluate usefulness and usability of Student List View (Q9-Q12), Question List View (Q13-Q14) and Behavior View (Q15-Q19). The third section of our questionnaire is to ask about some personal opinions on our system (Q20-Q21). The original sentences without the words in brackets are the positive statements at the right end of the scale points, while the sentences with words in the brackets are the negative statements at the left end.}
    \setlength{\aboverulesep}{0.5pt}
    \setlength{\belowrulesep}{0.5pt}
    \begin{tabular}{p{0.3cm}|p{7.4cm}}
        \toprule
         Q1& It is very easy~(difficult) to use. \\
         Q2& It is very easy~(difficult) to learn. \\
         Q3& I am very wiling~(unwilling) to use the system in the proctoring tasks. \\
         Q4& I am very~(not) confident on my selections using the system. \\ 
         Q5& I will~(will not) recommend the system to other TAs. \\ \hline
         Q6& The visual design is easy~(difficult) to understand. \\
         Q7& The visual design provides enough~(too little) information for me to find students who cheated. \\
         Q8& The visual design and interactions can~(cannot) help me find students who cheated. \\ \midrule
         Q9& It is very easy~(difficult) to find students of high risk in the Student List View. \\
         Q10& It is very easy~(difficult) to know the distribution of cheating types of a student. \\
         Q11& It is very easy~(difficult) to know the distribution of risk of different questions of a student. \\
         Q12& It is very easy~(difficult) to know the overall time used and time used for each question. \\ \hline
         Q13& It is very easy~(difficult) to select questions with the most suspected cases. \\
         Q14& It is very easy~(difficult) to know the distribution of cheating types of a student of each question. \\ \hline
         Q15& This view can~(cannot) help me better understand the suspected cases. \\
         Q16& It is very easy~(difficult) to know when there are suspected cases.\\
         Q17& It is very easy~(difficult) to know how a student moves his/her head and mouse in this view. \\
         Q18& It is very easy~(difficult) to compare the student's behavior with peers using the heatmap on the left side. \\
         Q19& It is very easy~(difficult) to compare the student's behavior with his/her behaviors on other questions using the heatmap on the right side.\\ \hline
         Q20& Do you have any suggestions for our system?\\
         Q21&  What do you think the advantages and disadvantages of our system are over the baseline method? \\ \bottomrule
        
    \end{tabular}
    \label{table:questions}
\end{table}



\begin{figure}[hbt!]
    \centering
    \includegraphics[width=\linewidth]{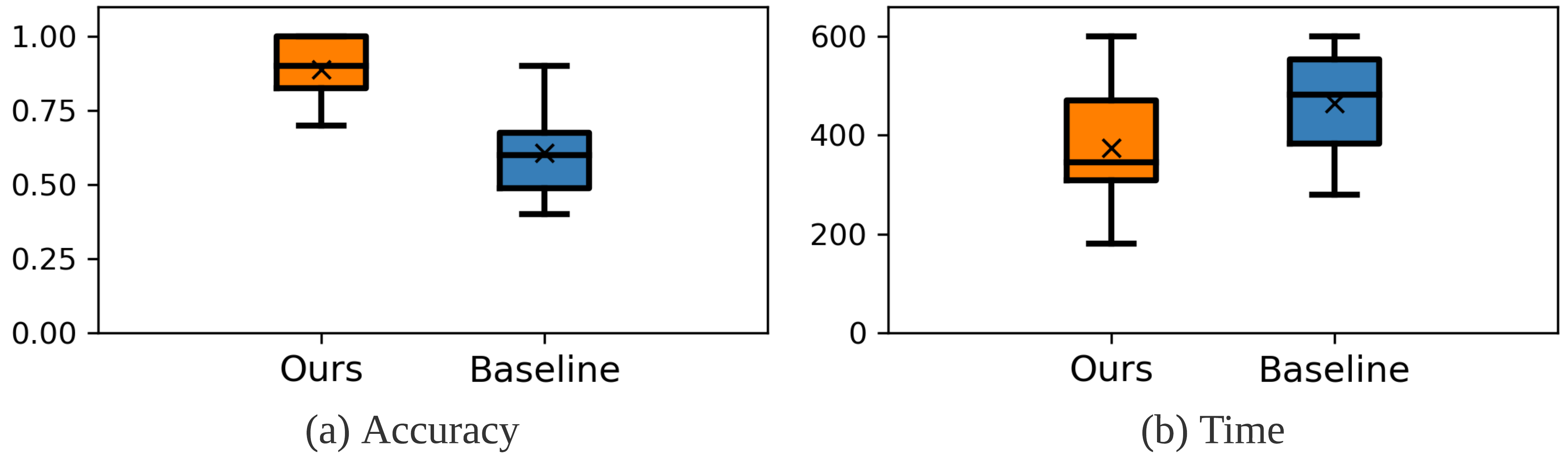}
    \vspace{-1em}
    \caption{Comparison of using our method and the baseline method in terms of (a) the accuracy of labeling cheating cases and (b) the total time cost for Task 1.}
    \Description[This figure shows comparison of using our method and the baseline method in terms of (a) accuracy and (b) time cost in Task 1.]{Our method outperforms the baseline method in both accuracy and time used to finish the task.}
    \label{fig:quantitative_result}
\end{figure}

\textbf{\haotian{Results of Task 1.}}
Quantitative results on the accuracy of labeling cheating cases and the total time for Task 1
are presented in Figure~\ref{fig:quantitative_result}. 
Our method outperforms the baseline method in both accuracy (our method: $0.888$, baseline: $0.606$) and time used to finish the task (our method: $374.375$ seconds, baseline: $463.938$ seconds). 
However,
the range of time used in our method is larger in Figure~\ref{fig:quantitative_result}(b). According to our observation in the user study, the reason for this phenomenon is that some participants would like to view all videos to make sure our system is reliable enough, while others preferred to rely more on our suspected case detection results and skip some videos without suspected cases.
To further demonstrate the efficiency and accuracy of our method, we conduct paired t-tests ($df = 15$) for every participant in terms of their accuracy and time cost in using both methods. The results of t-tests suggest that the difference between our method and the baseline approach are statistically significant in terms of both the accuracy ($t_{accuracy} = -7.028$, $p < 0.001$) and the time cost ($t_{time} = 2.301$, $p < 0.05$).

Since some experts mentioned that the false positive (a non-cheating case is labeled as a cheating one) rate and the false negative (a cheating case is labeled as a non-cheating one) rate are important in evaluating the cheating detection, we also report them in Table~\ref{table:false_positive_negative}. From the table, we can learn that both the average false positive rate and the average  false negative rate of our method are lower than those of the baseline. Furthermore, the standard deviations of the rates of our method are also lower than those of the baseline.

\begin{table}[hb!]
\centering
\caption{Comparison of our method and the baseline method in terms of the false positive rate and the false negative rate in Task 1. SD means standard deviation. The lower values are shown in bold.}
\Description[The table shows the comparison of our method and the baseline method in terms of the false positive rate and the false negative rate in Task 1.]{}
\vspace{-1em}
\label{table:false_positive_negative}
\footnotesize
\setlength{\aboverulesep}{0.5pt}
\setlength{\belowrulesep}{0.5pt}
\begin{tabular}{ccccc}

\toprule
                                     &         & Ours & Baseline \\ \midrule \midrule
\multirow{2}{*}{False positive rate} & Average &     \textbf{0.019}       &     0.313                   \\ \cmidrule{2-4}
                                     & SD     &       \textbf{0.054}     &            0.163            \\ \midrule
\multirow{2}{*}{False negative rate} & Average &       \textbf{0.210}     &       0.500                 \\ \cmidrule{2-4}
                                     & SD     &      \textbf{0.229}     &     0.239       \\ \bottomrule          
\end{tabular}
\end{table}

\begin{figure}[hbt!]
    \centering
    \includegraphics[width=\linewidth]{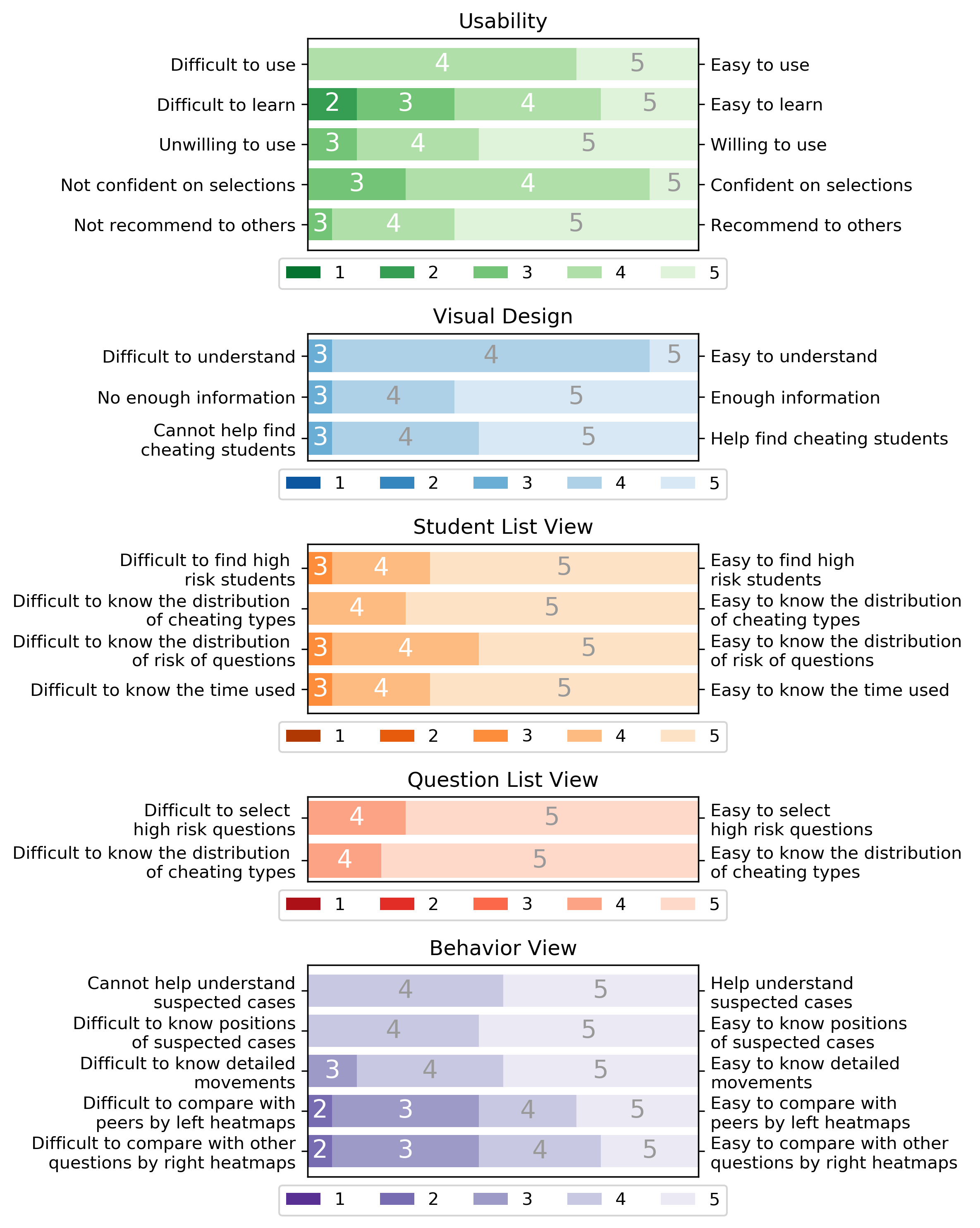}
    \vspace{-1em}
    \caption{The results of $Q1$-$Q19$ in our questionnaire. $1$-$5$ represents ``the most negative'' to ``the most positive''. \haotian{The number on each section shows the corresponding
    score. 
    }}
    \label{fig:qualitative_result}
    \Description[This figure shows the results of Q1-Q19 in our questionnaire in Task 2 of our user study.]{1-5 means from the most negative to the most positive. The negative statements in the questionnaire are on the left while the postive ones are on the right. In most of questions, our methods are highly rated by participants by giving us 4 or 5.}
    \vspace{-2em}
\end{figure}

\textbf{\haotian{Results of Task 2.}}
The results of our questionnaire are presented in Figure~\ref{fig:qualitative_result}. Overall, our system was highly rated by participants. They agreed that our system is quite convenient and efficient for proctoring. 
A participant commented that \textit{``Combining automated methods with visual analytics to facilitate the detection of abnormal cases vastly improves the efficiency and efficacy of proctoring online exams than simply watching the students' videos alone''}. 
Moreover, they found that Student List View is quite useful and it is easy to learn the information they needed, which serves as a complement to our results on accuracy and time cost mainly on other views. 
However, some participants worried that our system is not easy to learn and their main concerns are about the Behavior View. 
Several participants commented that the detailed behavior charts are not easy to understand at first glance, since the movements on X-axis and Y-axis are encoded individually.
However, after comparing our visual design with the alternative design in Figure~\ref{fig:alternative_design}, they agreed that our design can present information more clearly by reducing occlusion and understood that we need to strike a balance between intuitiveness and the clearness of information.
A participant suggested refining the legend and description of each column in the heatmaps to make them more understandable,
which has already been done in the final version of our system.


\subsection{Expert Interview}
We conducted in-depth interviews with four experts (P1, P2, P3, P5), who have been involved in our task analysis (Section~\ref{requirement_analysis}), through online or face-to-face meetings. Each interview started with a brief introduction to the visual encoding and interactions in our system. Then some cases were presented by us to further illustrate the usage of our system. After that, experts were invited to freely explore our system. They were encouraged to ask questions and comment on our system during the exploration. At the end of interviews, we asked them about their overall opinions about our system. Each interview lasted about 30-40 minutes and all the interviews were recorded with their permission. Due to personal reasons, P4 was not able to take an interview. 
Instead, we collected his feedback on our system through emails. 
Specifically, we sent him the link and the user guide to our system and invited him to freely explore the system. Then we asked about his opinion about our system's workflow and visual design.
Overall, our system is highly appreciated by all the experts. In this section, we summarize their feedback in 2 categories: suspected case detection and visualization system.

\textbf{Suspected Case Detection.}
The performance of our suspected case detection engine is considered quite satisfactory by the experts. 
All of them agreed that it is quite innovative 
and useful
to introduce the usage of mouse movements to online exam proctoring. They confirmed that mouse movements reveals rich information for cheating detection like leaving the current website. 
P2 and P5 thought our suspected case detection based on mouse movement data and the video recorded by a single webcam is able to help improve current online exam environment settings.
They pointed out that university students were required to set up multiple webcams from different angles in online exams by themselves, which led to non-standard online exam settings and made it hard for proctors to find cheating cases. 
In their opinion, our method provides a simple and unified online exam setting that is more convenient for both students and proctors.
P2 also commented that he would like to work with us to apply it in real online exams, since our mouse movement data collection module can be easily integrated into his learning management system. 
Considering the abnormal head movement detection, detection failure was a common concern of P1, P3 and P5. However, after knowing that the model we use can accurately estimate the head poses of a student with a face mask, they believed that our detection is quite reliable and helpful for finding abnormal head movements. 

The experts also provided some valuable suggestions for our suspected case detection.
P1 and P2 suggested that our detection engine should be able to handle more online exam settings, for example, setting up an extra webcam to record videos of body movement. We believed that this could be handled by adding extra detection modules to our engine. However, due to the limitations of our dataset, this part is left as our future work. P4 mentioned that we could try to extend the mouse movement collection plugin to the whole operating system and try to learn what the student did after leaving the web page.

\textbf{Visualization System.}
Overall, our visualization system is appreciated by all the experts due to its usefulness and usability. All the experts believed that our visualization system is quite intuitive and helpful in finding cheating cases by providing different levels of views. 
P1 liked the design of our system, and he said that the workflow is \textit{``well-designed and intuitive''}.
P2 commented that the overall UI design of our system is quite clear and professional. He quickly learned how to use our system and appreciated our interactions such as starting the video from a certain point by clicking on the detailed behavior charts~(\textbf{R5}).
He also thought that our system, especially the summary views (Student List View and Question List View) is able to help proctors locate cheating cases very fast and greatly reduce the workload of proctors to view the long and dull videos with great concentration~(\textbf{R1, R2}).  
P3 also commented \textit{``Currently we use a brute-force method to reviewing videos, the recommendation function provided by your system is very helpful''}.
By saying the ``recommendation'', he referred to displaying the risk level of students and questions in our Student List View and Question List View~(\textbf{R1, R2}). 
Besides, P3 confirmed that teachers can easily get enough information and find cheating cases from detailed behavior charts and heatmaps in our Behavior View~(\textbf{R3, R4}), but it might be overwhelming for student TAs.
He suggested 
providing
a simplified version for student TAs, which is left as our future work.

\section{Discussion}
\vspace{-2em}
\haotian{\subsection{Privacy Concerns}
Privacy concerns in live video streaming-based education have been recognized and discussed~\cite{chen2020LVS}, such as the unauthorized usage of video data by the third party and the unexpected exposure of living environment.
Similarly,
almost all online proctoring methods, including ours, also face similar concerns. For example, proctors need to record and check videos during and after online exams to identify possible cheating cases and guarantee online exam justice. 
However, to ensure a fair evaluation of students' performance, keeping the integrity of exams is crucial and is actually the responsibility of teachers~\cite{rogers2006faculty}.
Meanwhile, due to the lack of face-to-face interactions, it is much more challenging to
maintain the integrity of online exams than traditional classroom-based exams~\cite{king2014cheating, prince2009comparisons, richardson2013strengthening, rogers2006faculty}.
Thus, the proctors need to balance academic integrity and privacy concerns in online exams.
We believe that the method proposed in this paper can be regarded as an initial effort in striking a balance between academic integrity and privacy concerns.
It provides a more effective and efficient way for online exam proctoring. Meanwhile, strict measures are also needed to further address privacy concerns when it is used in real online exams.
For example, to avoid unauthorized data usage, the proctors need to set up a secure infrastructure and comprehensive regulations to store, use and delete the collected data appropriately.
Before the online exam, the detailed methods of data collection, processing, analysis and destruction should be revealed to students. 
Then students' consent to recording video and mouse movement data needs to be obtained. 
During the online exam, the usage of a virtual background can be permitted to hide bystanders and the living environment in videos. 
After the online exam, once all the cheating reviews are done, the data should be destroyed permanently.
}
\haotian{\subsection{Limitations of Collected Data}
As stated in Section~\ref{section:data_collection_setup}, we conducted a mock online exam to collect the mouse movement data, video data and cheating behavior labels, due to the lack of existing dataset and the difficulty of collecting cheating labels in real online exams. Though we had tried to mimic a real online exam environment and used compensation to 
encourage students to take the mock exam seriously,
our dataset may not
incorporate all possible cheating cases and all online exam settings.
First, the types of cheating cases in our dataset are limited.
In our mock online exams, since the benefit and risk of cheating were not as large as those in real online exams, participants mostly adopted the common means of cheating.
In real online exams, there are more advanced methods of cheating and some of them may be even harder to detect from their head and mouse movement, for example, using an earphone to listen to answers.
Our approach shows satisfactory performance of detecting common cheating behaviors like searching for answers through the Internet and using paper materials in our dataset, but its ability to deal with other cheating behaviors needs further evaluation.
Second, 
our mock online exam for cheating behavior data collection is conducted in the strictest closed-book setting, while real online exams may be conducted in other manners.
In different online exam settings, cheating behaviors can be different. 
For example, some open-book online exams may allow the usage of paper materials but prohibit the usage of search engines.
To facilitate the needs of different settings, we enable customized risk calculation to support adaptively filtering cheating behaviors, as mentioned in Section~\ref{section:suspicious_engine}.
As an initial exploration of using visual analytics techniques to facilitate the proctoring of online exams, we focus on the common cheating behaviors in closed-book online exams and leave the further research on other cheating behaviors as future work. 
To better address the limitations of data, more real-world datasets can be collected to extensively evaluate our system in the future.
}


\subsection{Real-time Proctoring}
\label{sec:real_time}
In our interviews, P1 and P2 commented that they would like to use our system for real-time proctoring, since teachers' intervention on cheating behaviors during the online exam is quite important. 
Currently, our method is used for reviewing videos after the online exam by proctors. However, it has the potential to be used for real-time proctoring if some issues can be addressed.
First, sufficient computational resources are required for real-time head pose estimation in our proctoring method.
In our approach, most of computational resources are consumed by extracting head poses from videos.
According to our experiment, an Nvidia Titan Xp GPU is needed for a student to extract his/her head poses in a real-time manner using the current model.
Since computational resources like GPUs are expensive, sometimes it may not be easy to provide enough resources to estimate a large number of students' head poses in online exams.
A possible method to mitigate the high demand of computational resources is to 
apply some
lightweight deep learning models like MobileNetV3~\cite{howard2019mobilenet} in our head pose estimation. 
\haotian{However, compared with more complicated deep learning models like ResNet-50 in Hopenet~\cite{ruiz2018hopenet}, 
the lightweight deep learning models may result in a performance drop.
Thus, users need to strike a balance between efficiency and accuracy when selecting models for real-time proctoring.}
Second, streaming mouse movement and video data need to be dealt with.
\haotian{In our method, we compute several statistical metrics of values (e.g., average yaw angle) using complete videos and mouse movement data. When switching to a real-time mode, these values can be computed by using sliding windows on the streaming data instead.}

\subsection{Generalizability and Scalability}
Our approach is designed for the proctoring of online exams. However, 
it is not limited to the proctoring of online exams. Instead, it 
can also be extended to
other applications.
For example, the coach of E-sports can conduct an analysis on a player's mouse movements to evaluate the effectiveness of each action and his/her response time. 
Also, the design of our Behavior View can facilitate the needs of presenting some other types of spatial-temporal data for in-depth analysis, for example, eye tracking data.

Also, the scalability of our system needs further discussion.
From the perspective of processing speed, abnormal head pose detection may take a long time or require much computational power when the number of students in the online exam is large. As we discussed in Section~\ref{sec:real_time}, a possible method to mitigate this issue is to apply lightweight deep learning models.
From the perspective of visual design, when an online exam is taken by lots of students or the number of suspected cases is too large, our system also suffers from scalability issues.
Though proctors can sort the Student List View for easy location of suspected students, it is still hard when the number of students is too large.
A possible solution is to apply hierarchical visualization methods to first group students and expand groups for individual inspection on demand.
Also, if the number of suspected cases is too large, the glyphs between two line charts in our Behavior View may have severe overlapping due to the limitation of screen size. A possible solution is to aggregate all suspected cases during a particular period (e.g., 30 seconds) as a pie chart.



\section{Conclusion and Future Directions}

Online exams have become increasingly popular for course instructors to assess the knowledge of students or other test takers.
However, it remains unclear on how to conveniently and effectively proctor online exams.
In this paper, we propose a visual analytics approach to achieve convenient, efficient and reliable online proctoring in this study. 
It consists of two major modules:
suspected case detection engine and visualization, which first processes students' videos and mouse movement data during the online exam and further visualizes them in three levels of details. 
We extensively evaluate our approach through three usage scenarios, a user study and in-depth interviews with experts. The results confirm the usefulness and effectiveness of our approach in enabling convenient, efficient and reliable proctoring for online exams.

In future work, we plan to improve our visual analytics approach for real-time proctoring and further evaluate our system in real-world online exams. Also, it will be interesting to explore how to apply visualization techniques to reduce cheating behaviors in online exams. 



\begin{acks}
This work is partially sponsored by Innovation and Technology Fund (ITF) with No. ITS/388/17FP. Yong Wang is the corresponding author. We would like to thank all participants in our mock online exam and the user study, all experts for their valuable opinions, all anonymous reviewers for their feedback, and Zezheng Feng and Jiakai Wang for their proofreading.
\end{acks}

\bibliographystyle{ACM-Reference-Format}
\bibliography{reference}

\appendix

\end{document}